\documentclass[%
 preprint, amsmath,amssymb,aps, prf]{revtex4-1}
\usepackage{float}
\usepackage{graphicx}
\usepackage{dcolumn}
\usepackage{bm}
\usepackage{color}
\usepackage{amsmath}

\begin{document}

\title{Compressible and immiscible fluids 
with arbitrary density ratios: diffuse interface theory}

\author{Fei Wang}
\email{fei.wang@kit.edu}

\affiliation{Institute of Nanotechnology (INT),
Karlsruhe Institute of Technology (KIT),\\
Hermann-von-Helmholtz-Platz 1, 76344 Eggenstein-Leopoldshafen, Germany}

\date{\today}

\begin{abstract}
For the water-air system,
the bulk density ratio is as high as about 1000;
no model can fully tackle such a high density ratio system.
In the Navier-Stokes and Euler equations, the density $\rho$ within
the water-air interface is assumed
to be a constant based on the Boussinesq 
approximation
namely $\rho (\mathrm{d}  \mathbf u/\mathrm{d} t)$,
which does not account for the true momentum evolution $\mathrm{d} (\rho \mathbf u)/\mathrm{d} t$ 
($\mathbf u$-fluid velocity).
Here, we present an alternative theory for the density evolution equations
of immiscible fluids in computational fluid dynamics, 
differing from the concept of Navier-Stokes and Euler equations.
Our derivation is built upon the physical principle of 
energy minimization from the aspect of thermodynamics.
The present results provide
a generalization of Bernoulli's principle for energy conservation
and a general formulation for the sound speed.
The present model can be applied for 
immiscible fluids with arbitrarily high density ratios, 
thereby, opening a new window for computational fluid
dynamics both for compressible and incompressible fluids.
\end{abstract}

\maketitle

\section{Introduction}

Immiscible fluids with a high density ratio are ubiquitous in our daily lives.
A typical example is the water-air system, where the density ratio is approximately 1000. 
However, addressing the density evolution in high-density-ratio systems remains an open question. 
In the literature, two typical approaches are used to treat density evolution at the water-air interface: (I) the Boussinesq approximation and (II) the interpolation method.

% \begin{itemize}
%  \item 
\textit{(I) Boussinesq approximation:} 
The first one is the Boussinesq approximation,
where the kinetic equation is written as 
\begin{equation}
 \rho \frac{\mathrm{d} \mathbf{u}}{\mathrm{d}t}=-\nabla p+\text{rest terms}.
 \label{eq:1-0}
\end{equation}
Here, $p$ is the hydrodynamic pressure and $\mathbf{u}$ represents the
fluid velocity.
In this approach, the density is assumed to be a constant everywhere at a fixed temperature,
i.e., $\rho=\text{constant}$. The limitation of the Boussinesq approximation 
is that the ``true'' momentum evolution  
\begin{equation}
\frac{\mathrm{d}  (\rho \mathbf{u})}{\mathrm{d} t}
\end{equation}
has not been taken into account
since the density evolution is overlooked.
Another limitation of the Boussinesq approximation
is the consideration of gravity and buoyancy which are actually 
caused by the difference in the density.
A constant density throughout the system cannot lead to gravity and buoyancy.
When gravity and buoyancy are much more pronounced than the inertial force,
a density difference term, $\Delta \rho=(\rho_w-\rho_a)\bold g$ 
is usually added to ``rest terms'' in Eq.~\eqref{eq:1-0}.
Here, $\rho_w$ and $\rho_a$ are the bulk densities of the water and air, respectively, 
and $\bold g$ is the gravity acceleration.

 \textit{(II) Interpolation method:}
A second approach is the interpolation of the density over the volume
concentration $\phi$ of water as~\cite{wu2024evolution,worner2012numerical} 
\begin{equation}
 \rho=\rho_w\phi+\rho_a(1-\phi).
  \label{eq:interpolation}
\end{equation}
This approach has
its advantage of computing gravity and buoyancy implicitly.
However, there are two limitations of the interpolation approach:
[II(A)]: Conflicts between incompressibility and diffusion.
[II(B)]: Uniform excess volume.

\begin{itemize}
 \item 
\textit{[II(A)]: Conflicts between incompressibility and diffusion}. 
On the one hand, the incompressible condition requires 
that 
\begin{equation}
 \frac{\mathrm{d} \rho}{\mathrm{d} t}=0.
\end{equation}
On the other hand, the total time derivative of the density according
to the interpolation approach
can be written as 
\begin{equation}
 \frac{\mathrm{d} \rho}{\mathrm{d} t}=(\rho_w-\rho_a)\frac{\mathrm{d} \phi}{\mathrm{d} t}\neq0.
\end{equation}
When diffusion takes place, i.e., $\mathrm{d} \phi/\mathrm{d} t\neq0$, 
we have $\mathrm{d} \rho/\mathrm{d} t\neq0$
if $\rho_w\neq\rho_a$, so that in the linear interpolation approach,
the incompressible condition is not compatible with diffusion
for two immiscible phases with distinct densities.
 \item 
\textit{[II(B)]: Uniform excess volume}.
According to the definition of density, we have
the following expression for the mixture density at any position
$\boldsymbol r$ in the considered domain $\Omega$ as 
\begin{equation}
 \rho=\frac{\rho_w v_w+\rho_a v_a}{V}=\rho_w \frac{v_w}{V}+\rho_a \frac{v_a}{V},
 \label{eq:3}
\end{equation}
where $V\subset\Omega$ is a volume element including  
sufficient amount of fluid particles for statistic thermodynamics,
and expressed as 
$V=v_w+v_a+v_e$; $v_e$ depicts the excess volume 
and $v_i$ ($i=w,\ a$) is the volume of the $i$ species. 
By assuming a zero excess volume, i.e., $v_e=0$, we have
\begin{equation}
 V=v_w+v_a.
 \label{eq:6-00}
\end{equation}
Substituting Eq.~\eqref{eq:6-00}
into Eq.~\eqref{eq:3} and using the definition of volume concentration
\begin{equation}
\phi_i=v_i/V,
 \label{eq:6-000}
\end{equation}
The linear interpolation, Eq.~\eqref{eq:interpolation} is replicated.
Suffice it to say, 
the linear interpolation according to Eq.~\eqref{eq:interpolation} 
indicates a zero excess volume of mixing
at every position of the considered domain $\Omega$. 
Any other kinds of interpolations define
a certain excess volume, which is also uniform everywhere in the domain $\Omega$.
The assumption of uniform excess volume (either zero or non-zero)
deviates from the physical fact
that the local excess volume $v_e$
is affected by the local pressure, which should be space dependent;
the excess volume at each position in the domain
should have its own characteristics, which will be considered in the current work. 
A detailed discussion of our concept regarding the excess volume is presented in Section III.
\end{itemize}
% \end{itemize}

Note that only for the uniform zero excess volume,
we have the following constraint
\begin{equation}
 \phi_w+\phi_a=1.
 \label{eq:8-00}
\end{equation}
When $V\neq v_a+v_w$, the above constraint, Eq.~\eqref{eq:8-00} generally does not hold.
In summary, to solve the high-density-ratio problem, two evolution equations are required
\begin{itemize}
 \item one is for the volume concentration;
 \item the other one is for the excess volume or the molar volume of the mixture.
\end{itemize}

In this work, we explore the high-density ratio problem through the physical principle of energy minimization. We propose a generalized theoretical framework for the evolution equations of fluid dynamics and derive a novel kinetic equation governing density dynamics at the fluid-fluid interface.
The resulting density evolution equation yields three key contributions:
(A):  A universal formulation for sound speed;
(B):  A generalized equation of state for fluids;
(C): A generalization of Bernoulli’s law, extending the conservation of potential and kinetic energy. 
For each contribution, a case study and application is presented in Section VII.

The rest of the paper is structured as follows.
In Section~\ref{sec:2}, we provide a brief overview of two high-density ratio models from literature, 
clarify the distinction between the total time derivative and the mass conservative derivative, and examine the validity of the conventional continuity equation.
Section~\ref{sec:3} outlines the setup of our study and introduces a new concept of volume concentration.
In Section~\ref{sec:6}, we present a dissipation-conservation principle that establishes a connection between mass conservation and energy dissipation.
Section~\ref{sec:7} applies this principle to derive the classical fluid dynamics equations for uniform systems.
Building on these results, Section~\ref{sec:8} extends the derivation to non-uniform systems, derives the free energy dissipation, and formulates a novel kinetic equation for density.
In Section~\ref{sec:case}, we demonstrate the application of our theory by deriving the sound speed, the equation of state (EOS) for fluids, and the density profile across a water-air interface.
Finally, Section~\ref{sec:14} concludes the paper with a summary of our findings.

\section{Two models for high density ratio system in literature}
\label{sec:2}
\subsection{Ding-Spelt-Shu approach}
The Navier-Stokes equation for incompressible fluids reads~\cite{ding2007diffuse}
\begin{equation}
 \rho \frac{\mathrm{d}\mathbf{u}}{\mathrm{d}t}=\rho\bigg( \frac{\partial \mathbf{u}}{\partial t}+\mathbf{u}\cdot \nabla \mathbf{u}\bigg)=-\nabla p+\nabla\cdot [\eta(\nabla \mathbf{u}+\nabla \mathbf{u}^T)]+\bold f_s,
 \label{eq:7-0}
\end{equation}
where $\eta$ is the dynamic viscosity
and $\bold f_s$ depicts the surface tension force.
The notation $\nabla \mathbf{u}^T$ means $(\nabla \mathbf{u})^T$.

The total time derivative of linear momentum is expressed as 
\begin{align}
\notag
 \frac{\mathrm{d} (\rho \mathbf{u})}{\mathrm{d}t}&
 =\frac{\partial (\rho \mathbf{u})}{\partial t}+
 \nabla\cdot (\rho \mathbf{u}\otimes \mathbf{u})\\
 &=\frac{\partial \rho}{\partial t} \mathbf{u}+\rho\frac{\partial \mathbf{u}}{\partial t} + (\nabla\cdot \rho \mathbf{u})\mathbf{u}+\rho \mathbf{u}\cdot \nabla \mathbf{u} \notag \\
 &=\rho\bigg(\frac{\partial \mathbf{u}}{\partial t} +\mathbf{u}\cdot \nabla \mathbf{u}\bigg)+ \bigg[\frac{\partial \rho}{\partial t}
 +\nabla\cdot (\rho \mathbf{ u})\bigg]\mathbf{u}.
 \label{eq:13-000}
\end{align}
Here, the following vector calculus
has been used 
\begin{equation}
 \nabla\cdot (\bold a\otimes \bold b)
 = (\nabla \cdot \bold a)\bold b+\bold a\cdot \nabla \bold b. 
\end{equation}
So that an additional time evolution equation 
for the density is required to iterate the time evolution
of the momentum, as shown in Eq.~\eqref{eq:13-000}.

According to the linear interpolation, 
as adopted in Ref.~\cite{ding2007diffuse}, the total time derivative 
of the density reads
\begin{equation}
 \frac{\mathrm{d}\rho}{\mathrm{d}t}=(\rho_w-\rho_a)\frac{\mathrm{d}\phi}{\mathrm{d}t}=(\rho_a-\rho_w)\nabla\cdot \mathbf{j},
 \label{eq:12}
\end{equation}
where the diffusion equation with diffusion flux $\mathbf{j}$ has been used
\begin{equation}
 \frac{\mathrm{d}\phi}{\mathrm{d}t}=\frac{\partial \phi}{\partial t}+\mathbf{u}\cdot \nabla\phi=-\nabla \cdot\mathbf{j}.
\end{equation}

In Ding-Spelt-Shu approach~\cite{ding2007diffuse}, a term $(\mathrm{d}\rho/\mathrm{d}t)\mathbf{u}$
is added to both sides of Eq.~\eqref{eq:7-0},
leading to 
\begin{equation}
 \frac{\mathrm{d}(\rho\mathbf{ u})}{\mathrm{d}t}=-\nabla p+\nabla\cdot [\eta(\nabla \mathbf{u}+\nabla \mathbf{u}^T)]+\bold f_s+(\rho_a-\rho_w)(\nabla\cdot \mathbf{j})\mathbf{u}.
 \label{eq:17-00}
\end{equation}
When removing  $(\mathrm{d}\rho/\mathrm{d}t)\mathbf{u}$ from both sides of Eq.~\eqref{eq:17-00},
the Navier-Stokes equation, Eq.~\eqref{eq:7-0} is replicated.

\subsection{Abels-Garcke-Grün approach}

The Abels-Garcke-Grün approach~\cite{abels2012thermodynamically} differs from the Ding-Spelt-Shu method.
In Abels-Garcke-Grün approach,
 the momentum equation is assumed to be written as 
\begin{equation}
 \frac{\mathrm{d}(\rho\mathbf{ u})}{\mathrm{d}t}=-\nabla p+\nabla\cdot \bold T+\bold f_s,
 \label{eq:14}
\end{equation}
with an unknown stress tensor $\bold T$.
By moving the time derivative of the density to the right hand side,
Eq.~\eqref{eq:14} is rewritten as 
\begin{equation}
 \rho\frac{ \mathrm{d}\mathbf{ u}}{\mathrm{d}t}=-\nabla p+\nabla\cdot \bold T+\bold f_s-\frac{\mathrm{d}\rho}{\mathrm{d}t}\mathbf{u}.
 \label{eq:15}
\end{equation}
Substituting Eq.~\eqref{eq:12}
into Eq.~\eqref{eq:15} leads to 
\begin{equation}
 \rho\frac{ \mathrm{d}\mathbf{ u}}{\mathrm{d}t}=-\nabla p+\nabla\cdot \bold T+\bold f_s-(\rho_a-\rho_w)(\nabla\cdot \mathbf{j})\mathbf{u},
 \label{eq:16}
\end{equation}
It is apparent that the last term is not objective that a coordinate 
transformation, such Newtonian transformation, will change the formulation of the equation~\cite{alt2009entropy}.
This non-objective problem is solved by Abels, Garcke, and Grün
via the following method:
By using the vector calculus
$\nabla \cdot (\mathbf{j}\otimes\mathbf{u} )=(\nabla\cdot \mathbf{j})\mathbf{u}+\mathbf{j}\cdot \nabla \mathbf u$,
the last term in Eq.~\eqref{eq:16} is rewritten as 
\begin{equation}
 (\nabla\cdot \mathbf{j})\mathbf{u}=\nabla \cdot ( \mathbf{j}\otimes\mathbf{u} )-\mathbf{j}\cdot \nabla \mathbf u.
\end{equation}
Abels, Garcke, and Grün define a new stress tensor by combining $\nabla \cdot ( \mathbf{j}\otimes\mathbf{u} )$ with $\nabla\cdot \bold T$, namely
\begin{equation}
 \tilde {\bold T}= \bold T+\mathbf{j}\otimes\mathbf{u}:=\eta(\nabla \mathbf{u} 
 + \nabla \mathbf{u}^T).
\end{equation}
In this way, the modified Navier-Stokes equation in the work of Abels-Garcke-Grün is expressed as 
\begin{equation}
 \rho\frac{\mathrm{d} \mathbf{u}}{\mathrm{d}t}=
 -\nabla p+\nabla\cdot \eta (\nabla \mathbf{u}+\nabla \mathbf{u}^T)
 +(\rho_a-\rho_w)\mathbf{j}\cdot \nabla \mathbf{u}.
\end{equation}
The last term is proportional to the gradient of the velocity, which is now objective.
However, the physical meaning for the definition of the new stress tensor $\tilde {\bold T}$ is unclear. Specially, 
it is unclear why a symmetric stress tensor
$\tilde {\bold T}$ is obtained by combining a unknown stress tensor $\bold T$
with an asymmetric stress tensor $\mathbf{j}\otimes\mathbf{u}$
for arbitrary velocity $\mathbf{u}$  and arbitrary diffusion flux $\mathbf{j}$.

Other existing literatures are based either on the continuity equation for the density evolution~\cite{tryggvason2001front,puckett1997high,fedkiw1999non} (which will be discussed in Section II C and Section VI D) 
or on the linear interpolation of the density~\cite{fakhari2017improved,shukla2010interface} according to Eq.~\eqref{eq:interpolation}.

\subsection{Total time derivative, mass conservation derivative, and validity of the continuity equation}
\label{sec:4}
We often encounter two types of derivatives when deriving the kinetic equation in fluid dynamics:
(I) the total time derivative and
(II) the conservative derivative.
 We clarify the difference between these two derivatives, as follows.
The total time derivative reads
\begin{equation}
 \frac{\mathrm{d} \rho}{\mathrm{d} t}= \frac{\partial \rho}{\partial t}+\frac{\partial \rho}{\partial x}\frac{\mathrm{d} x}{\mathrm{d} t}+
 \frac{\partial \rho}{\partial y}\frac{\mathrm{d} y}{\mathrm{d} t}+
 \frac{\partial \rho}{\partial z}\frac{\mathrm{d} z}{\mathrm{d} t}=
 \frac{\partial \rho}{\partial t}+\mathbf{u}\cdot \nabla \rho.
\end{equation}
The mass conservative derivative reads~\cite{tryggvason2001front}
\begin{align}
\notag
 \frac{\mathrm{D}\rho}{\mathrm{D}t}
 &=\frac{\partial \rho}{\partial t}+\nabla\cdot (\rho \mathbf u)\\
 &=\frac{\mathrm{d} \rho}{\mathrm{d} t}+\rho\nabla\cdot\mathbf u.
\end{align}
When there is no diffusion, i.e., $\mathrm{D} \rho/\mathrm{D} t=0$, we obtain
\begin{equation}
 \frac{\mathrm{d} \rho}{\mathrm{d} t}=-\rho \nabla\cdot \mathbf{u}.
 \label{eq:24}
\end{equation}
According to Eq.~\eqref{eq:24},
the divergence free of the fluid velocity $\nabla\cdot\mathbf{u} $
is equivalent to the incompressible condition of 
$\mathrm{d} \rho/\mathrm{d} t=0$, but contradicts with 
the case with diffusion, where $\mathrm{d} \rho/\mathrm{d} t\neq0$.

\begin{figure}[h!]
 \centering
 \includegraphics[width=0.65\linewidth]{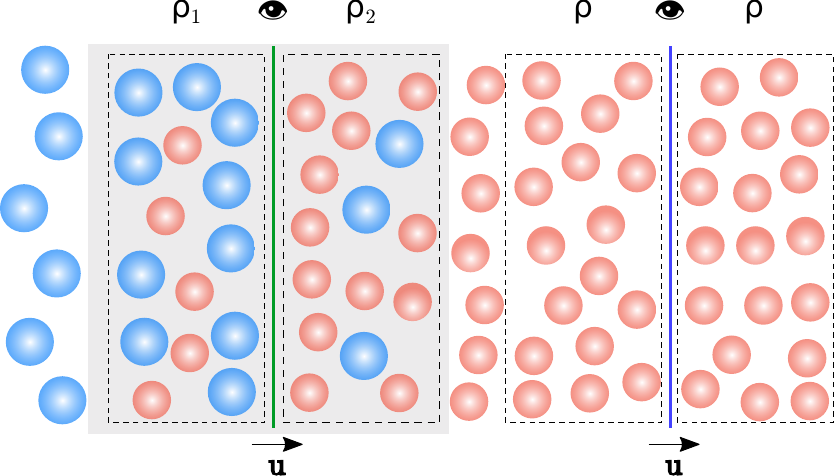}
 \caption{Sketch for the validity of the convection flux in a water-air system.
The water-air interface consists of water (blue) and gas (orange) molecules.
Within the water-air interface, the density is not uniform.
For example, when measuring the convection flux across the reference
line (green), the density at the left and right hand side of the reference line is different. 
}
 \label{fig:1-0}
\end{figure}
We note that, in general, the so-called mass conservative derivative should not be used to derive the kinetic equation. The key issue is that mass conservation has not been connected to the physical principle of dissipation, 
as will be demonstrated below. 
Furthermore, the mass-conservative derivative poses difficulties for deriving the continuity equation, as illustrated below and discussed in detail in Appendix D.
% \section{Validity of the continuity equation}
% \label{sec:5}
% Next, we demonstrate the limitations of the conventional continuity equation for describing density evolution in the presence of an interface.
In the absence of diffusion, the convective mass flux $\mathbf{j}_c$ is commonly expressed in the literature as
\begin{equation}
 \mathbf{j}_c=\rho \mathbf{u},
 \label{eq:25-0}
\end{equation}
which has been often adopted to derive the so-called continuity equation
\begin{equation}
 \frac{\partial \rho}{\partial t}+\nabla \cdot (\rho \mathbf u)=0.
 \label{eq:25-1}
\end{equation}
We emphasize that Eq.~\eqref{eq:25-0} 
can only be used to formulate the mass flux caused by convection in the absence of spatial density variations. However, when the density varies in space, this mass flux expression, Eq.~\eqref{eq:25-0}, becomes invalid.
As shown in Fig.~\ref{fig:1-0}, we consider a system consisting 
of water (blue) and gas (orange) molecules.
In the bulk of the gas phase, 
the convection flux across the reference line (highlighted in blue) with a fluid
velocity $\mathbf{u}$
can be readily expressed as $\rho \mathbf{u}$.
Within the water-air interface, the density varies from $\rho_1$
to $\rho_2$ from the left to the right side of the reference line (green line).
In this case, a unique density $\rho$ cannot be used to measure the convection flux.
Suffice it to say,
 the conventional continuity equation, Eq.~\eqref{eq:25-1}
 cannot be applied in regions where an interface is present.
 The fundamental problems are two-fold:
 \begin{itemize}
  \item [(P1)] When there is a spatial variation in density, the relationship between mass conservation and energy dissipation becomes unclear in the continuity equation, Eq.~\eqref{eq:25-1}.
  
  \item [(P2)]  When there is a spatial variation in density,
  how are the molecules convected from the region of $\rho_1$
 to the region of $\rho_2$, 
 and how do the former one occupy the latter region?
 Are there any changes of the excess volume as molecules 
 move from the $\rho_1$ region to the $\rho_2$ region?
 \end{itemize}

\section{Setup, definition and excess volume concept}
\label{sec:3}

 We consider an isothermal, closed system composed of two immiscible phases, denoted by $\alpha$ and $\beta$. Each phase contains $K \in \mathbb{Z}$ chemical components within a domain $\Omega$. To illustrate how our formulation differs from existing models, we take a water–gas system as an example. The bulk densities of the water and gas phases are denoted by $\rho_w$ and $\rho_a$, respectively.
We assume that the water–gas interface has a finite thickness, within which water and gas molecules coexist in a mixed region. A representative volume element  $V$ is considered, containing a statistically sufficient number of fluid particles in the interfacial region to evaluate the local fluid density as
\begin{equation}
 \rho=\frac{m_w+m_a}{V}=\frac{\rho_w v_w+\rho_a v_a}{V},
\end{equation}
where $v_w$ and $v_a$ denote the partial volumes of the water and gas components, respectively, and $m_w$ and $m_a$ represent the masses of water and gas contained within the volume element.
When $V = v_w + v_a$, corresponding to an ideal mixture with zero excess volume throughout the system, the overall density $\rho$ can be expressed as a linear interpolation of the volume fractions, defined as $\phi_i = v_i / V$, as introduced earlier.
The assumption of zero excess volume in every representative element across the water–gas interface is commonly adopted in the literature and is often referred to as the incompressible fluid approximation. However, enforcing exactly zero excess volume implies that the density varies monotonically from the bulk water phase to the bulk gas phase. This contradicts recent findings indicating that the density profile across the water–air interface exhibits a non-monotonic behavior~\cite{pezzotti20182d,creazzo2024water}.

Instead of assuming zero excess volume throughout the system, we propose a more general approach (Fig. 2a) in which 
\begin{equation}
V=v_w+v_a+v_e,
\end{equation}
where $v_e \neq 0$ denotes the excess volume, which exhibits spatial variation as a result of local pressure changes. Instead of explicitly solving for $v_e$, we redistribute it between $v_w$ and $v_a$ by an appropriate allocation scheme, resulting in two subregions with volumes $v_w^\prime$ and $v_a^\prime$. Consequently, the total volume satisfies $V = v_w^\prime + v_a^\prime$ (Fig. 2b).
This redistribution of excess volume implies that the local partial densities of water and gas must vary accordingly, since the individual masses of the two components within the volume element $V$ remain constant.
We define the volume composition including the information of excess volume as
\begin{equation}
 \phi_w=\frac{v_w^\prime}{V}\quad \text{and} \quad  \phi_a=\frac{v_a^\prime}{V}.
 \label{eq:10-0}
\end{equation}
Therefore, the fluid density within the interface 
is calculated as
\begin{equation}
 \rho=\rho_w^\prime\frac{v_w^\prime}{V} +\rho_a^\prime\frac{v_a^\prime}{V}=\rho_w^\prime\phi +\rho_a^\prime(1-\phi).
\label{eq:30-1}
\end{equation}
Note that both $\rho$ and $\phi$ encode the information associated with the excess volume.
The partial densities $\rho_w^\prime$ and $\rho_a^\prime$ are implicitly influenced by the excess volume and are therefore no longer constant, as they vary with the redistribution of the excess volume. Consequently, to fully characterize the state of the system, two evolution equations are required: one governing the volume composition $\phi$, and the other governing the  density $\rho$.

The classical treatment of incompressible flow corresponds to a special case in which the excess volume vanishes both in the bulk phases and across the interface. Under this assumption, it is sufficient to describe the system using a single variable—either the density $\rho$ or the volume composition $\phi$.
A representative example employing $\phi$ is the van der Waals formulation of free energy and its associated diffuse-interface framework~\cite{wohrwag2018ternary,ashour2023phase,bestehorn2021faraday,jacqmin1999calculation,anderson1998diffuse,yue2004diffuse,lowengrub1998quasi}, in which a single phase order parameter $\phi$ is used to characterize the system’s state and evolution.
This well-established approach~\cite{ridl2018lattice} is commonly referred to as the phase-field model or Cahn–Hilliard model. Another classical example is Euler’s formulation for density evolution, where the standard continuity equation, Eq.~\eqref{eq:25-1}, is adopted. For incompressible flow, Euler’s formulation reduces to the condition that the convective velocity field is divergence-free.
It is important to note that, for incompressible fluids, the van der Waals and Euler formulations cannot be applied simultaneously, as doing so would lead to an overdetermined system.

\begin{figure}[h!]
 \centering
 \includegraphics[width=0.6\linewidth]{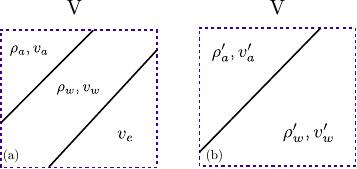}
 \caption{The volume concentration within the volume element $V$.
 (a) Conventional definition.
 The volume element $V$ is divided
 into the volume of air ($v_a$) and the volume of water ($v_w$), 
 plus the excess volume $v_e$.
 In this definition, the densities $\rho_a$ and $\rho_w$ are constants.
 (b) Current concept. The volume element $V$ is divided
 into the volume of air ($v_a^\prime$) and the volume of water ($v_w^\prime$).
 We have distributed the excess volume into the 
 volumes of $v_a^\prime$ and  $v_w^\prime$. 
 By this way, the densities  $\rho_a^\prime$ and $\rho_w^\prime$ are affected by the local excess volume 
 and  no more constants.
 }
 \label{fig-1}
\end{figure}

The present work develops a generalized theoretical framework for systems with nonzero excess volume within the fluid–fluid interface, particularly for cases involving large density ratios. The objective is to extend both the classical van der Waals and Euler formulations by incorporating interfacial compressibility. Here, compressibility refers to the variation of density across the interface arising from mixing with a finite excess volume. Even when the excess volume is nonzero, 
if $\rho_w^\prime \approx \rho_a^\prime = \rho^*$, the single-variable van der Waals formulation remains adequate to represent the compressibility, since $\rho \approx \rho^*$ according to Eq.~\eqref{eq:30-1}. Either the variation of $ \rho^*$  or $\phi$
can then describe the compressibility.
However, when $\rho_w^\prime \gg \rho_a^\prime$, the van der Waals formulation loses validity in the absence of a prescribed equation of state (EOS) and must therefore be generalized to accurately describe the evolution of the density field.

The above concept can be readily extended to the more general case of a multicomponent system, for which the following equality holds:
\begin{equation}
 \phi_i=\frac{v_i^\prime}{V}\quad \text{and} \quad\sum_i v_i^\prime=V,\quad \text{and} \quad\sum_{i=1}^K \phi_i=1.
\end{equation}
The boundary of the domain $\Omega$ is denoted by $\Gamma$.
For a binary system, we use the notation $\phi_1=\phi$, 
$\phi_2=1-\phi$.
The density of the mixture at any position $\boldsymbol r\in \Omega$
is denoted by $\rho(\boldsymbol r)$.

\section{Dissipation principle}
\label{sec:6}
In this section, we present a dissipation-conservation theorem, 
which will be used to derive the kinetic equation in the following sections.

The dissipation-conservation theorem
describes the energy dissipation 
for the field $u$ and the associated potential $\Phi$.
 It establishes a connection between the first law of thermodynamics (conservation) 
with the second law of thermodynamics (dissipation).
The energy is formulated as a function of $u$ as 
\begin{equation}
 E=\int_\Omega e(u) \mathrm{d} \Omega.
 \label{eq:30-00}
\end{equation}
The time derivative reads
\begin{equation}
 \frac{\mathrm{d} E}{\mathrm{d} t}=\int_\Omega \frac{\mathrm{d} e}{\mathrm{d} u} \frac{\mathrm{d} u}{\mathrm{d} t}\mathrm{d} \Omega=\int_\Omega \Phi \frac{\mathrm{d} u}{\mathrm{d} t}\mathrm{d} \Omega,
 \label{eq:5-20}
\end{equation}
where we have defined a potential $\Phi=\mathrm{d} e/\mathrm{d} u$.
From Eq.~\eqref{eq:5-20}, we obtain the conserved evolution of $u$ as 
\begin{equation}
 \frac{\mathrm{d} u}{\mathrm{d} t}=\nabla \cdot \tau_0 \nabla \Phi=\nabla \cdot \tau_0 \nabla \frac{\mathrm{d} e}{\mathrm{d} u},
 \label{eq:5-3}
\end{equation}
which is consistent with standard theory of diffusion equation.
The coefficient $\tau_0$ is a positive constant.
Substituting Eq.~\eqref{eq:5-3} into Eq.~\eqref{eq:5-20},
integrating by parts, and using the no-flux boundary condition, we obtain
\begin{equation}
 \frac{\mathrm{d} E}{\mathrm{d} t}=\int_\Omega \frac{\mathrm{d} e}{\mathrm{d} u}\frac{\mathrm{d} u}{\mathrm{d} t}  \mathrm{d} \Omega 
 =
 \int_\Omega \frac{\mathrm{d} e}{\mathrm{d} u} \nabla \cdot \tau_0 \nabla \Phi \mathrm{d} \Omega 
 =-\int_\Omega \tau_0(\nabla\Phi)^2  \mathrm{d} \Omega\leq0. 
 \label{eq:5-4}
\end{equation}

In summary, an energy dissipation in the form of Eq.~\eqref{eq:5-20}
simultaneously implies
a conservation equation in the form of  Eq.~\eqref{eq:5-3},
and vice versa.

From Eq.~\eqref{eq:5-3} and Eq.~\eqref{eq:5-4},  we observe that the system energy is non-increasing as long as $\tau_0\geq 0$.
The physical interpretation of $\tau_0$ can be understood as follows. Applying the chain rule, we rewrite Eq.~\eqref{eq:5-3} in the form of a diffusion equation:
\begin{equation}
 \frac{\mathrm{d} u}{\mathrm{d} t}=\nabla \cdot \bigg(\tau_0 \frac{\mathrm{d} ^2e}{\mathrm{d} u^2} \nabla u\bigg)=\nabla \cdot (\mathcal D \nabla u).
\end{equation}
Thus, the coefficient $\tau_0$ establishes a connection between the purely dynamic parameter $\mathcal  D$  (independent of conservative energy)
and the energy-dependent term  $\frac{\mathrm{d} ^2e}{\mathrm{d} u^2}$ via the relation $\tau_0= \mathcal D /\frac{\mathrm{d} ^2e}{\mathrm{d} u^2}$.
% In other words, $\tau_0$ represents the ratio of non-conservative to conservative energy contributions to the kinetics of the system.
% In specific cases, the physical meaning of $\tau_0$ will be demonstrated as follows:
% \begin{itemize}
%  \item In the case of mass diffusion for free energy dissipation, $\tau_0= \mathcal  D /(\partial^2 f/\partial c^2)$, where $\mathcal  D$ is the mass diffusion coefficient.  This aligns with the Cahn-Hilliard theory (see Section V).
%  \item In the case of convection for kinetic energy dissipation, $\tau_0=\mathcal  D/(\partial^2 K/\partial u^2)=\eta/\rho$ (see Section VI B), where $\mathcal  D=\eta/\rho$ is the dynamic viscosity. This is consistent with Newton's viscous dissipation theory (see Section VI B).
% \end{itemize}

The dissipation equation for non-conserved variables
can  be written
in the steepest descendent approach:
\begin{equation}
 \frac{\mathrm{d} u}{\mathrm{d} t}=-\tau_0 u.
 \end{equation}

\section{Uniform system: Derivation of Euler's equation and Newton's law}
\label{sec:7}
In this section, we apply the dissipation-conservation principle to a uniform system 
to derive the conventional fluid dynamics equations.
We suppose that the total energy 
of the system consists of the free energy $\mathcal  F$, the pressure energy $\mathcal  P$,
and the kinetic energy $\mathcal  K$
\begin{equation}
 \mathcal E= \mathcal  F+ \mathcal  P+ \mathcal  K.
\end{equation}
The free energy, pressure energy, and kinetic energy
are integrated over the energy density as 
\begin{align}
 \mathcal  F&=\int_\Omega f(\phi, T) \mathrm{d} \Omega,\\
 \mathcal  P&=\int_\Omega p \mathrm{d} \Omega,\\
 \mathcal  K&=\int_\Omega k \mathrm{d} \Omega =\int_\Omega \rho \mathbf{u}\cdot  \mathbf{u} \mathrm{d} \Omega.
\end{align}
The free energy density depends on the volume concentration $\phi$ and temperature $T$,
and may be expressed as $f=\theta-Ts$, where $\theta$ denotes the internal energy
and $s$ represents the entropy.

The total time derivative 
of the total energy reads
\begin{equation}
 \frac{\mathrm{d}  \mathcal  E}{\mathrm{d} t}=\int_\Omega \frac{\partial f}{\partial T}\frac{\mathrm{d} T}{\mathrm{d} t}+ \frac{\partial f}{\partial \phi}\frac{\mathrm{d} \phi}{\mathrm{d} t} + \frac{\mathrm{d} p}{\mathrm{d} t} 
 + \frac{\mathrm{d}  (\rho \mathbf{u})}{\mathrm{d} t}\cdot \mathbf{u}+ \rho\mathbf{u}\cdot\frac{\mathrm{d}   \mathbf{u}}{\mathrm{d} t} \mathrm{d} \Omega.
  \label{eq:4}
\end{equation}
Next, by using the chain rule for the space dependent pressure $p(\mathbf r)$,
we rewrite the total time derivative of the pressure as 
\begin{equation}
 \frac{\mathrm{d}  p}{\mathrm{d} t}=\frac{\mathrm{d} p}{\mathrm{d} \mathbf{r}}\cdot \frac{\mathrm{d}  \mathbf r}{\mathrm{d} t}=\nabla p\cdot \mathbf{u}.
 \label{eq:5-2}
\end{equation}
Substituting Eq.~\eqref{eq:5-2} into Eq.~\eqref{eq:4}, we obtain
\begin{equation}
  \frac{\mathrm{d} \mathcal  E}{\mathrm{d} t}=\int_\Omega \frac{\partial f}{\partial T}\frac{\mathrm{d} T}{\mathrm{d} t}+ \frac{\partial f}{\partial \phi}\frac{\mathrm{d} \phi}{\mathrm{d} t}  + 
 \bigg[ \nabla p+\frac{\mathrm{d}  (\rho \mathbf{u})}{\mathrm{d} t}\bigg]\cdot \mathbf{u}+ \rho\mathbf{u}\cdot\frac{\mathrm{d}   \mathbf{u}}{\mathrm{d} t} \mathrm{d} \Omega.
 \label{eq:42-0}
\end{equation}

Applying the dissipation principle stated in Section~\ref{sec:6}
to the first part of Eq.~\eqref{eq:42-0},
we obtain Fourier’s heat equation for the temperature as
\begin{equation}
 \frac{\mathrm{d} T}{\mathrm{d} t}=\nabla \cdot M_T\nabla \frac{\partial f}{\partial T}= \nabla \cdot M_T(\partial^2f/\partial T^2)\nabla T=\nabla\cdot D_T\nabla T,
\end{equation}
where the thermal diffusivity $D_T$ is defined in terms of the mobility $M_T$ as $D_T=M_T(\partial ^2f/\partial T^2)$.
Since the thermal diffusivity generally exceeds the mass diffusivity by at least three orders of magnitude, the temperature evolution equation is consequently neglected in the following derivation. For instance, at room temperature, the thermal diffusivity of water is approximately $0.14\times10^{-6}$ m$^2$/s,
whereas its mass diffusivity is only $2.3\times10^{-9}$ m$^2$/s.

Applying the dissipation principle 
to the second part of Eq.~\eqref{eq:42-0},
we obtain a conserved time evolution equation for $\phi$,
which is the Fick's diffusion equation, as 
\begin{equation}
 \frac{\mathrm{d} \phi}{\mathrm{d} t}=\nabla \cdot M\nabla \frac{\partial f}{\partial \phi}= \nabla \cdot M(\partial ^2f/\partial \phi^2)\nabla\phi= \nabla\cdot D\nabla \phi,
\end{equation}
where the mass diffusivity $D$ is defined in terms of the mobility $M$ as $D=M(\partial ^2f/\partial \phi^2)$.

When considering that
there is no velocity dissipation, e.g., inviscid fluids, we obtain the momentum balance equation according to the second term in Eq.~\eqref{eq:42-0}
\begin{equation}
 \frac{\mathrm{d} (\rho \mathbf{u})}{\mathrm{d} t}=-\nabla p,
\end{equation}
which depicts the exchange of the kinetic energy with the potential energy
via the work done by the pressure $\nabla p\cdot \mathbf{u}$
and the work done by the inertial force, $\mathbf  f_\rho\cdot \mathbf{u}=\rho \mathbf{u}\cdot \mathbf{u}$.

When the density is assumed to remain constant over time,
\begin{equation}
 \frac{\mathrm{d} \rho}{\mathrm{d} t}=0,
\end{equation}
the momentum balance equation is written as 
\begin{equation}
 \rho \frac{\mathrm{d} \mathbf{u}}{\mathrm{d} t}=-\nabla p,
\end{equation}
which is nothing but the Euler's equation in fluid dynamics.
An integration over the whole domain, we obtain the well-known Newton's law
\begin{equation}
 \mathbf F=m \mathbf a,
\end{equation}
where the mass $m$, the acceleration $\mathbf a$, 
and the force $\mathbf F$ are defined as 
\begin{align}
        m   =\int_V\rho \mathrm{d} V,\ 
 \mathbf a =\frac{\mathrm{d}  \mathbf{u}}{\mathrm{d} t},\
 \mathbf F  =\int_V -\nabla p\mathrm{d} V.
\end{align}

\section{Non-uniform system: System energy with an interface}
\label{sec:8}
In this section, we apply the dissipation-conservation principle
to non-uniform systems and derive the corresponding kinetic equations.

\subsection{Energy formulation and dissipation}

We propose that the total energy of the system 
consists of free energy $\mathcal F$, 
 pressure  energy $\mathcal P$,
macroscopic kinetic energy $\mathcal K$, and wall free energy $ \mathcal W$, reading
\begin{align}
\mathcal E =\mathcal F+ \mathcal P + \mathcal K +\mathcal W.
\label{eq:50}
\end{align}
This is a standard approach in the literature, except for the inclusion of the pressure term,  $\mathcal P$.
The motivation for introducing $\mathcal P$ will be presented below.

The integration over the local free energy density $e(\boldsymbol\phi, \nabla\boldsymbol\phi)$
leads to
\begin{align}
&\mathcal F(\boldsymbol\phi, \nabla\boldsymbol\phi)=\int_\Omega e(\boldsymbol\phi, \nabla\boldsymbol\phi) \mathrm{d} \Omega.
%\\
%&e(\boldsymbol\phi, \nabla\boldsymbol\phi)=f(\boldsymbol \phi)
%+ \sum_{i=1}^K\frac{1}{2}\kappa_i(\nabla\phi_i)^2,
\label{eq:5}
\end{align}
% where $f(\boldsymbol \phi)$ depicts
% the bulk free energy density and $\kappa_i$ represents the gradient energy
% % coefficient which is 
% related to the interfacial tension of the $\alpha$-$\beta$ interface, 
% $\sigma_{\alpha\beta}$.
By decomposing the entropy $s$ into thermal entropy $s_T$, 
component-mixing entropy $s_\phi$, and velocity entropy $s_u$, we write the free energy density as 
\begin{equation}
 e=\theta-Ts=\theta-T(s_T+s_\phi+s_u)=e_0-Ts_u,
 \end{equation}
 where $\theta$ represents the internal energy.
 The physical meaning of the entropies $s_\phi$ and $s_u$
 is as follows:
 \begin{itemize}
  \item The component mixing entropy $s_\phi$ accounts for the mixing excess volume in the diffusion equation, thereby leading to compressibility.
  \item The velocity entropy $s_u$ is associated with viscous effects, resulting in velocity dissipation.
  \end{itemize}
 The free energy $e_0=f-T (s_T+s_\phi)$ does not contain the contribution 
 of the velocity entropy. Here, $f=f_0+ \sum_{i=1}^K\kappa_i(\nabla \phi_i)^2$ depicts
the van der Waals-Cahn free energy~\cite{cahn1958free,rowlinson1979translation} and $\kappa_i$ represents the gradient energy
coefficient.
It should be noted that the entropy increase of 
 $s_\phi$ and $s_u$ does not necessarily result in heat release or a temperature change of 
 the same magnitude as that associated with thermal entropy. 
 For instance, mass diffusion can lead to an increase in 
 $s_\phi$
 in an approximately isothermal manner, since the mass diffusivity is much smaller than the thermal diffusivity.
 Similarly, dissipation of fluid velocity can increase 
 $s_u$  nearly isothermally, as the kinematic viscosity of water
 is one magnitude smaller than its thermal diffusivity.
For these physical reasons, we assume isothermal conditions over the time scale of water–mass diffusion in the present work.

% \begin{figure}[h!]
% \centering
%  \includegraphics[width=0.7\linewidth]{pressure1.pdf}
%  \caption{Origin of the pressure energy for uniform and nonuniform
%  systems. For a uniform system (a), the pressure
%  at any position is the same, which is taken as a reference
%  value. In a non-uniform system with density variation (b),
%  the pressure is generally non-uniform due to the asymmetric
%  collisions of different sized molecules. Blue and red spheres
%  depict two molecules or atoms with distinct sizes, leading to
%  a variation of the density in space. The arrows sketch the
%  thermal motion of the molecules/atoms.
%  }
%  \label{fig}
% \end{figure}

The $\mathcal K$ term denotes the kinetic energy of the fluid, which is formulated as
 \begin{equation}
 \mathcal K=\int_\Omega \boldsymbol \zeta \cdot \mathbf u \mathrm{d}\Omega,
\end{equation}
where $\boldsymbol \zeta$ is the linear
momentum density expressed  as 
$\boldsymbol \zeta= \rho \mathbf{u}$.

The $\mathcal P=\int_\Omega p \mathrm{d}\Omega$ term, 
often overlooked previously and introduced here, describes the pressure energy.
In existing works,
the space gradient of pressure is usually employed as the driving force
for the fluid motion in the Navier-Stokes equation
by using the isothermal Gibbs-Duhem relation
\begin{equation}
 -\mathrm{d} p + \sum_{i=1}^K \phi_i\mathrm{d} \mu_i=0,
 \label{eq:57-0000}
\end{equation}
where $\mu_i$ is the chemical potential of species $i$, defined as $\mu_i=\delta e/\delta \phi_i$.
Here, $\delta$ denotes the functional derivative.
However, we note that there are two major limitations
when using the gradient of pressure $\nabla p$ based on   $\mu_i=\delta e/\delta \phi_i$
and Eq.~\eqref{eq:57-0000}
to derive the momentum evolution.
\begin{itemize}
 \item First, the classical Gibbs–Duhem relation, Eq.~\eqref{eq:57-0000}, is derived under a quasi-static assumption, in which the velocity-dependent entropy $s_u$ and the associated kinetic energy  are not taken into account. Consequently,
when $\nabla p$ is used as given in Eq.~\eqref{eq:57-0000}, the role of viscosity in free-energy dissipation or in maximizing the velocity entropy is not clearly defined. One cannot simply introduce a viscous term into the momentum evolution in an ad hoc manner without accounting for its fundamental connection to entropy.
 \item Second, the space derivative $\nabla$ and the functional derivative $\delta$ (or $\mathrm{d}$) are not  mathematically compatible in most existing works except few references~\cite{ridl2018lattice} (see Appendix B). 
%Since the derivative of the free energy with respect to volume concentration is expressed as a functional derivative,
%i.e., $\delta e= \sum_{i=1}^K\mu_i \delta \phi_i$, all derivatives appearing in the first law of thermodynamics should likewise be written in terms of functional derivatives. 
%For example, $-p\delta v+\sum_{i=1}^K\mu_i \delta \phi_i$. 
% Note that, when the chemical potential is defined as $\mu_i=\partial e/\partial \phi_i-\nabla \cdot (\partial e/\partial \nabla \phi_i) $~\cite{ledesma2014lattice}, $\nabla \mu_i\neq \delta \mu_i/\delta x$.
% However, if we would modify the definition for the chemical potential as:
% $\mu_i=\mathrm{d} e/\mathrm{d}\phi_i= \partial e/\partial \phi_i + (\partial e/\partial \nabla \phi_i)\cdot \nabla\phi_i$,
% as used  in Ref.~\cite{ridl2018lattice},
% the $\delta$ operator is consisent with the $\nabla$ operator.
% The key problem here is that only after applying the integration by parts 
% to the term $ (\partial e/\partial \nabla \phi_i)\cdot \nabla\phi_i$ on the whole domain $\Omega$,
% we obtain the mostly used result, $-\nabla \cdot (\partial e/\partial \nabla \phi_i)$
% in literature; the negative sign is caused by the integration by parts.
\end{itemize}

To solve these two limitations, we derive a modified Gibbs-Duhem relation as follows.
The total derivative of the kinetic energy reads, $\delta k= \mathbf{u}\cdot \delta(\rho \mathbf{u})+ \rho \mathbf{u} \cdot\delta \mathbf{u}$.
The first part $ \mathbf{u}\cdot \delta(\rho \mathbf{u})$ is the inertial work which can be absorbed into the first law of thermodynamics in addition to
the chemical work $\mu_i  \delta\phi_i$ and mechanical work $p \delta v$.
As will be shown in Section~VIB, this term, $ \mathbf{u}\cdot \delta(\rho \mathbf{u})$ combing with pressure
results in the momentum evolution.
The second part $\rho \mathbf{u} \cdot\delta \mathbf{u}$ is left into the modified isothermal Gibbs-Duhem relation
\begin{equation}
 -\delta p + \sum_{i=1}^K \phi_i\delta \mu_i+\rho \mathbf{u} \cdot\delta \mathbf{u}=0,
 \label{eq:59-0000}
\end{equation}
as the spirit of Gibbs-Duhem relation is that 
the total derivative of the energy minus the work associated with the first law of thermodynamics is zero.
As $\rho \mathbf{u} \cdot\delta \mathbf{u} =-T\delta s_u$ (see Section~VIB),  we integrate Eq.~\eqref{eq:59-0000} 
and obtain a general definition of the pressure as (see Appendix A) 
\begin{equation}
 -p=e-\sum_{i=1}^K \mu_i \phi_i+Ts_u=e_0-\sum_{i=1}^K \mu_i \phi_i.
  \label{eq:60-00}
\end{equation}
This result provides a clear understanding of the pressure and is fully consistent with previous works~\cite{chen1998lattice}.
It must be noted that the pressure $p$ in  Eq.~\eqref{eq:60-00} does not include the velocity-dependent entropy,
as $e+Ts_u$ results in a $s_u$ independent energy $e_0$.

The $\mathcal W$ term in Eq.~\eqref{eq:50} represents the wall free energy,
which is integrated over the wall free energy density $f_w$ on the boundary
$\Gamma$ as $\mathcal W=\int_\Gamma f_w \mathrm{d}\Gamma$.
The  wall free energy density $f_w$
is responsible for the wetting phenomenon,
which has been comprehensively discussed elsewhere~\cite{wang2024wetting,wang2023thermodynamically,wang2021wetting}.

In the following, we derive the equilibrium condition and the kinetic equations
based on the principle of energy minimization,
namely 
\begin{equation*}
\frac{\mathrm{d}  \mathcal E}{\mathrm{d}  t}=\dot {\mathcal E}\leq0.
\end{equation*}
Next, we evaluate the total time derivative 
of the free energy $\mathcal{F}$, 
the pressure energy $\mathcal{P}$,
and the kinetic energy $\mathcal{ K}$. 
By using the chain rule and noting that
the functional $\mathcal F$ has two arguments, $\boldsymbol\phi$ and $\nabla\boldsymbol\phi$,
the total time derivative of the potential free energy $\mathcal F$ reads
\begin{equation}
\dot{\mathcal F}(\boldsymbol\phi,\nabla\boldsymbol\phi )
 =\int_\Omega \sum_{i=1}^K\frac{\partial e}{\partial \phi_i}\frac{\mathrm{d} \phi_i}{\mathrm{d}  t}
 + \frac{\partial e}{\partial \nabla\phi_i}\cdot \frac{\mathrm{d} \nabla \phi_i}{\mathrm{d}  t}
 \mathrm{d}\Omega.
\end{equation}
By using the vector calculus $\nabla \cdot (\nabla\phi_i \otimes  \nabla\phi_i)
 =\nabla^2 \phi_i \nabla\phi_i + (\nabla\phi_i\cdot \nabla) \nabla\phi_i$,
 the definition of materials derivative $\dot {\nabla \phi_i}=\partial_t\nabla\phi_i+\mathbf{u}\cdot \nabla\nabla \phi_i$,
and integration by parts with no-flux boundary condition (more details can be found in~\cite{wang2023thermodynamically,zhang2024multi}), we obtain the following equations
\begin{align}
\label{eq_14}
&\dot{ \mathcal F}
 =\int_\Omega \sum_{i=1}^K
 \mu_i  \dot{\phi}_i
 + (\nabla\cdot \underline{\underline{\boldsymbol \Theta}})\cdot \mathbf{u}
   \mathrm{d}\Omega;\\
 & \mu_i= \frac{\delta  e }{\delta \phi_i}=\frac{\partial e}{ \partial \phi_i}
 -\nabla \cdot \frac{\partial e}{ \partial \nabla\phi_i};
  \label{eq_14_0}\\
 &\underline{\underline{\boldsymbol \Theta}}=\sum_{i} \frac{\partial e}{\partial \nabla\phi_i}\otimes \nabla \phi_i.
   \label{eq_14_4}
\end{align}
The first part of Eq.~\eqref{eq_14} defines a generalized formulation
for the chemical potential, as stated in Eq.~\eqref{eq_14_0}.
At thermodynamic equilibrium, the chemical potential 
is a constant value throughout the system.
In a special case of a zero chemical potential, i.e., $\mu_i=0$, Eq.~\eqref{eq_14_0} is equivalent
to the Euler-Lagrange equation and replicates the Young-Laplace equation;
Eq.~\eqref{eq_14_0} also provides a physical 
interpretation for the variational derivative in mathematics.
The second part of Eq.~\eqref{eq_14} defines a stress tensor $\underline{\underline{\boldsymbol \Theta}}$,
which is responsible for the transformation of the potential free energy
into the macroscopic kinetic energy~\cite{jacqmin1999calculation}.
This stress tensor is consistent with the Korteweg stress 
by evaluating the derivative, $\partial_{\nabla\phi_i} e  =\kappa_i\nabla\phi_i$.

Noteworthily, a uniform chemical potential throughout the system
cannot well define the thermodynamic equilibrium; 
the pressure evolution should be considered as well.
An evaluation of the time derivative for the pressure energy
and the kinetic energy
leads to
\begin{align}
\label{eq9}
 &\dot {\mathcal P}= \int_\Omega\frac{\mathrm{d}  p(\mathbf{r})}{\mathrm{d}  t}\mathrm{d}  \Omega =\int_\Omega \nabla p\cdot \dot{ \mathbf{r}} \mathrm{d} \Omega
 =\int_\Omega \nabla\cdot (p \mathbf{I}) \cdot \mathbf{u} \mathrm{d} \Omega,\\
 & \dot {\mathcal K}
 =\int_\Omega ( \dot{\boldsymbol \zeta} \cdot \mathbf{u} + 
 \boldsymbol \zeta\cdot \dot{\mathbf{u}}) \mathrm{d} \Omega,
 \label{eq10}
\end{align}
where $\mathbf I$
is an identity tensor.
For the derivation of Eq.~\eqref{eq9} and Eq.~\eqref{eq10},
we have used the chain rule and the definition of the velocity, $\dot{ \mathbf{r}}=\mathbf{u}$ without considering relativity effect~\cite{landau2013fluid}.
Note that $\mathrm{d} p/\mathrm{d} t=\partial p/\partial t+\mathbf{u}\cdot \nabla p$.
For fluids with velocity much less than sound speed, $\partial p/\partial t\approx0$.
Summarizing Eq.~\eqref{eq_14}, Eq.~\eqref{eq_14_0}, Eq.~\eqref{eq_14_4}, Eq.~\eqref{eq9}, and Eq.~\eqref{eq10}
and collecting all the $\cdot \mathbf{u}$ terms,
we obtain the total time derivative of the system energy $\mathcal E$ as 
\begin{align}
\label{eq13}
&\dot{\mathcal E}=\int_\Omega   \sum_{i=1}^K \underbrace{\mu_i \dot{\phi_i}}_{\text{I}}+ 
\underbrace{[\nabla\cdot (\underline{\underline{\boldsymbol \Theta}}+p\mathbf{I}) + \dot{\boldsymbol \zeta}]}_{\text{II}} \cdot \mathbf{u}
+ \underbrace{\boldsymbol \zeta\cdot \dot{\mathbf{u}}}_{\text{III}} \mathrm{d}\Omega.
\end{align}
% The physical interpretation for
% the three parts of the energy evolution in Eq.~\eqref{eq13} is sketched in Fig.~\ref{fig2}.
% 
% \begin{figure}[h!]
%  \centering
%  \includegraphics[width=0.7\linewidth]{pressure2.pdf}
%  \caption{The energy dissipation mechanisms in Eq.~\eqref{eq13}.
%  I: The diffusion via the exchange of the molecules and vacancies
%  leads to the dissipation of the free energy $\mathcal F$.
%  II: The gradient of the free energy $\mathcal F$
%  and the associated pressure $\mathcal P$ 
%  result in an increase in the momentum $\boldsymbol\zeta=\rho \mathbf{u}$.
%  III: The difference in the velocity, density, and viscosity
%  in distinct subdomains (dashed rectangle)
%  gives rise to the dissipation of the velocity  $\mathbf u$
%  caused by the `spring' drag effect;
%  the spring symbol denote the interaction between the atoms/molecules in different subdomains.}
%  \label{fig2}
% \end{figure}

Note that, at isothermal condition,  the total energy is conserved when defining 
$\Lambda=\mathcal E+Ts_\phi+Ts_u$,
where the entropy energies of component and velocity have been added.
Within this framework, we write the time derivative of the  total energy as 
\begin{align}
\frac{\mathrm{d}  \Lambda}{\mathrm{d}  t}=\int_\Omega   \sum_{i=1}^K \bigg(\mu_i \frac{\mathrm{d}  \phi_i}{\mathrm{d} t}+T\frac{\mathrm{d}  s_{\phi_i}}{\mathrm{d}  t}\bigg) +
[\nabla\cdot (\underline{\underline{\boldsymbol \Theta}}+p\mathbf{I}) + \dot{\boldsymbol \zeta}] \cdot \mathbf{u}
+ \bigg( \boldsymbol \zeta\cdot \frac{\mathrm{d}  \mathbf{u}}{\mathrm{d}  t}+T\frac{\mathrm{d}  s_{u}}{\mathrm{d}  t}\bigg)  \mathrm{d} \Omega=0.
\end{align}

\subsection{Kinetic equations} 
According to the dissipation-conservation theorem, the first part in Eq.~\eqref{eq13}, I, defines 
the generalized diffusion-convection equation in a non-uniform system as 
\begin{align}
\label{eq_14_1}
 \dot{\phi_i}&=\frac{\mathrm{d} \phi_i}{\mathrm{d} t}=\frac{\partial \phi_i}{\partial t}+\mathbf{u}\cdot\nabla\phi_i =\nabla \cdot \varsigma_i\nabla \mu_i.
\end{align}
The coefficient $\varsigma_i$ is the mobility.
The physical meaning is that the diffusion leads to an 
increase in the mixing entropy $s_{\phi}$.

The second part in Eq.~\eqref{eq13}, II, reveals the momentum conservation
\begin{equation}
 \dot{\boldsymbol \zeta}=\frac{\mathrm{d}  (\rho \mathbf{u})}{\mathrm{d} t}= -\nabla\cdot (\underline{\underline{\boldsymbol \Theta}}+p\mathbf{I}).  
 \label{eq_14_3}
\end{equation}
The conservative force at the right hand side is responsible for the exchange 
of the potential energy $\mathcal F+ \mathcal P$
with the  kinetic energy $\mathcal K$.
This result is nothing but momentum balance.
Here, a more clear physical meaning 
of the force is presented; 
the force term  consists of
the pressure $p$ and the stress tensor $\underline{\underline{\boldsymbol \Theta}}$.
The stress tensor is due to the gradient term in the potential energy $ \mathcal F$,
while the pressure is contributed by the associated free energy.
Note that a viscous dissipation term cannot be added to the momentum conservation equation, 
as the pressure term  does not contain the information of the velocity entropy [see Eq.~\eqref{eq:60-00}].

The only term left in the time evolution of system free energy, III,
is $\int_\Omega \boldsymbol \zeta\cdot \dot{\mathbf{u}} \mathrm{d} \Omega$.
By applying the same dissipation principle as done for the scalar field  
in Eq.~\eqref{eq_14_1},
we obtain the dissipation equation
for the velocity as 
\begin{equation}
 \dot {\mathbf u}=\nabla \cdot \tau_1(\nabla  \boldsymbol\zeta+\nabla\boldsymbol\zeta^T)
 +\nabla \cdot \tau_2 (\nabla\cdot \boldsymbol\zeta)\mathbf{I}.
 \label{eq17}
\end{equation}
Unlike the dissipation rule for the scalar variable $\phi_i$ with the associated potential
$\nabla \mu_i$ [see Eq.~\eqref{eq_14_1}],
three terms $\nabla \boldsymbol\zeta$,  $\nabla \boldsymbol\zeta^T$, and $(\nabla\cdot \boldsymbol\zeta)\mathbf{I}$
have to be taken into account for the dissipation of a
vector $\mathbf{u}$. 
The physical meaning of the velocity dissipation
is the increase of the velocity entropy $s_u$.
Within the framework of entropy,
the third term, III, in  Eq.~\eqref{eq13} can be interpreted as 
\begin{equation}
 \boldsymbol \zeta\cdot \frac{ \mathrm{d} \mathbf{u}}{ \mathrm{d}  t}=-T\frac{\mathrm{d}  s_u}{\mathrm{d}   t}.
 \label{eq:70-999}
\end{equation}
The left hand side is responsible 
for the minimization of the kinetic energy,
which is equivalent to the right hand side
for the  maximization of the velocity entropy.
% Eq.~\eqref{eq:70-999} gives rise to
% a new understanding of Newton's law for 
% the acceleration.
% Suffice it to say, the acceleration $\bold a= \mathrm{d} \mathbf{u}/\mathrm{d}  t$
% results from the entropy increase.
% This concept is consisent with the work of Verlinde~\cite{verlinde2011origin}.

The right hand side of Eq.~\eqref{eq17} is equivalent to Newton 
viscous stress tensor if we define 
the mobilities $\tau_1$ and $\tau_2$ as 
\begin{equation}
 \tau_1=\eta/\rho^2,\  \tau_2=\nu/\rho^2,
 \end{equation}
where $\eta$ and $\nu$ are the shear and bulk viscosities~\cite{tanaka2000viscoelastic}, respectively.

The dissipation equation, Eq.~\eqref{eq17}, is essentially a diffusion equation applied to the vector $\mathbf u$, as evident from the following comparison:
\begin{align}
&\nabla\cdot \tau_1 \nabla \boldsymbol\zeta=\nabla\cdot \tau_1 \nabla (\partial  k/\partial \mathbf u);\ \boldsymbol\zeta=\partial k/\partial \mathbf u;\\
&\nabla\cdot \varsigma \nabla \mu=\nabla\cdot \varsigma \nabla  (\partial e/\partial \mathbf \phi);\
 \mu=\partial  e/\partial \phi.
\end{align}
The linear momentum, $\boldsymbol\zeta$, plays a role analogous to that of the chemical potential, $\mu$.

A further question is that if the velocity
is a conserved variable or not.
If not, the dissipation has to be modified by the 
non-conserved form as 
\begin{equation}
 \frac{\mathrm{d}  \mathbf{u}}{\mathrm{d} t}=-\tau_0 \boldsymbol\zeta,
\end{equation}
The non-conserved form 
is known as the Allen-Cahn type dissipation
following the gradient descent path, being consistent with the Langevin equation.

\subsection{Density evolution equation and the generalization of Bernoulli's law}
\label{sec:11}
Eq.~\eqref{eq_14_3} subtracting Eq.~\eqref{eq17}
with  the relation 
$ \dot{\boldsymbol\zeta}=\dot{\rho}\mathbf{u}+\rho\dot{\mathbf{u}}$,
we obtain the temporal equation
for the density as 
\begin{equation}
 \mathbf u\dot \rho
 =-\nabla \cdot [(p\mathbf{I}+\underline{\underline{\boldsymbol \Theta}})]
 -\rho\nabla \cdot [ \tau_1(\nabla  \boldsymbol\zeta+\nabla\boldsymbol\zeta^T)+\tau_2 (\nabla\cdot \boldsymbol\zeta)\mathbf{I}] .
 \label{eq23}
\end{equation}
As shown below, Eq.~\eqref{eq23} provides a generalized form of Bernoulli’s law, a general definition of the sound speed, and a generalized equation of state (EOS).

% For inviscid flow, there is no energy dissipation for kinetic energy, i.e.,
% $\tau_1=\tau_2\approx0$.
% In this  case, Eq.~\eqref{eq23}
% is rewritten as 
% \begin{equation}
%  \mathbf u \frac{d\rho}{dt}
%  =-\nabla\cdot (p\mathbf{I}+\underline{\underline{\boldsymbol \Theta}}).
% \end{equation}
% By using the definition of the velocity and pressure gradient in one dimension,
% $u=dr/dt$, and $\nabla p=dp/dr$,
% we obtain the relation between velocity, pressure, and density as 
% \begin{equation}
% u^2=\bigg(\frac{d p}{d \rho}\bigg)_s,
% \label{eq:28-1}
% \end{equation}
% where the stress tensor has been overlooked.
% The result stated by Eq.~\eqref{eq:28-1} is fully consistent
% with the definition of sound speed.
% The condition for this definition is that there is no 
% energy dissipation, namely, isentropic, as indicated by the  subscription $s$
% denoting the isentropic condition.
% This result coincides with the theory of sound speed~\cite{landau2013fluid}.

Considering inviscid fluids via setting $\tau_1=\tau_2\approx0$,
and the steady state,
$\dot{\rho}=\mathbf{u}\cdot \nabla \rho$
where we set $\partial_t\rho=0$, we rewrite Eq.~\eqref{eq23}
as 
\begin{equation}
 \mathbf{u}(\mathbf u\cdot  \nabla \rho)
 =-[\nabla \cdot (p\mathbf{I}+\underline{\underline{\boldsymbol \Theta}})].
 \label{eq25}
\end{equation}
By considering a one-dimensional setup
and using the relation $\mathrm{d} (\rho u^2)=u^2 \mathrm{d} \rho+ 2\rho u \mathrm{d}u$,
Eq.~\eqref{eq25} is further simplified
as 
\begin{equation}
 \mathrm{d}  (p+\rho u^2)=2\rho u\mathrm{d} u.
\end{equation}
This result is nothing but a generalized Bernoulli's equation.
When there is no change of the velocity entropy,  
namely, $\rho u \mathrm{d}u=0$,
we have 
\begin{equation}
 p+\rho u^2=\text{constant},
\end{equation}
which is Bernoulli's principle.
However, when velocity entropy increases, $\mathrm{d} u\neq0$,  the associated energy 
is changed as well;
in this case, the classic Bernoulli's equation
loses its validity.

\subsection{The problem of the conventional kinetic energy formulation}
Conventional formulation for  the kinetic energy is  $\tilde{\mathcal K}=\int_\Omega \rho \mathbf{u}^2 \mathrm{d} \Omega$.
 In this case, the time derivative of the kinetic energy reads
\begin{equation}
 \frac{\mathrm{d} {\tilde{\mathcal K}} }{\mathrm{d} t}=\int_\Omega \bigg(
 \frac{\mathrm{d} \rho}{\mathrm{d} t}
 \mathbf{u}^2+\rho\mathbf{u}\cdot \frac{\mathrm{d} \mathbf{u}}{\mathrm{d} t} \bigg)\mathrm{d} \Omega.
 \label{eq:26}
\end{equation}
A noteworthy shortcoming 
of Eq.~\eqref{eq:26}
is the missing information of the momentum balance as
well as the transformation between the potential and kinetic energies.
Moreover, the first part in Eq.~\eqref{eq:26} requires a prior information 
about the time evolution of the density, which is unknown for 
two immiscible phases within the interface,
as the excess volume within the interface is unknown.
It is important to note that the classical continuity equation, $\partial_t\rho+\nabla\cdot (\rho \mathbf{u})=0$, cannot be applied here for density evolution (see Appendix D).

\subsection{Replication of the Navier-Stokes equations}
\label{sec:12}
In this section, we show that in a special case, the present theory is consisent with Navier-Stokes equations.
Only in this section, we will consider the factor $1/2$
for the kinetic energy $(1/2)\rho \mathbf{u}\cdot\mathbf{u} $.

The momentum balance equation with the factor $1/2$ reads
\begin{equation}
\frac{1}{2}\frac{\mathrm{d} (\rho \mathbf{u}) }{\mathrm{d} t} =\frac{1}{2}\rho \frac{\mathrm{d}  \mathbf{u} }{\mathrm{d} t}+ \frac{1}{2}\mathbf{u}\frac{\mathrm{d} \rho}{\mathrm{d} t}=-\nabla p.
  \label{eq:47-0}
\end{equation}
For incompressible fluids $\mathrm{d} \rho/\mathrm{d} t=0$, Eq.~\eqref{eq:47-0}
reduces to 
\begin{equation}
 \frac{1}{2}\rho \frac{\mathrm{d}  \mathbf{u} }{\mathrm{d} t}=-\nabla p.
 \label{eq:48}
\end{equation}
Multiplying both sides of the dissipation equation, Eq.~\eqref{eq17}
by $\rho$, the dissipation equation reads [a factor $1/2$ has been added due to
the kinetic energy formulation, $(1/2)\rho \mathbf{u}\cdot\mathbf{u}$ ]
\begin{equation}
 \frac{1}{2} \rho \frac{\mathrm{d}  \mathbf{u}}{\mathrm{d} t}=\rho \{\nabla\cdot \tau_1[ \nabla(\rho \mathbf{u}) + \nabla(\rho \mathbf{u})^T]+\nabla \cdot \tau_2 \nabla\cdot(\rho\mathbf{u})\mathbf{I}\}=\nabla\cdot \eta( \nabla \mathbf{u} + \nabla \mathbf{u}^T)
 +\nabla\cdot [\nu(\nabla\cdot \mathbf{u})\mathbf{I}].
 \label{eq:49}
\end{equation}

Summing Eq.~\eqref{eq:48}
with Eq.~\eqref{eq:49}, 
we obtain the classic Navier-Stokes equation for incompressible fluids as
\begin{equation}
 \rho \frac{\mathrm{d}  \mathbf{u}}{\mathrm{d} t}=-\nabla p +\nabla\cdot \eta( \nabla \mathbf{u} + \nabla \mathbf{u}^T)+\nabla\cdot \nu(\nabla\cdot \mathbf{u})\mathbf{I}.
\end{equation}

We stress that viscous dissipation inherently contradicts the incompressibility condition.
When viscous dissipation occurs, density cannot remain constant, as dissipative processes introduce entropy changes that modify the density.
However, under dissipation-free conditions (e.g., isentropic flow), density can reasonably be treated as constant.
In this way, the Navier-Stokes equation has the conflicts
between  viscous dissipation and invariance of density (incompressibility).
In fact, Eq.~\eqref{eq:48} cannot be summed with Eq.~\eqref{eq:49} 
because:
\begin{itemize}
 \item  Eq.~\eqref{eq:48} is based on the incompressible condition 
that the density does not vary with time; 
\item  Eq.~\eqref{eq:49} depicts a kind of viscous dissipation
due to entropy increase.
Strictly speaking, when viscous dissipation occurs,
the density cannot be a constant.
\end{itemize}
When there is viscous dissipation,
we must use Eq.~\eqref{eq:47-0} with the total time derivative of
the momentum, rather than Eq.~\eqref{eq:48}.
To solve the total time derivative of
the momentum, a key difficulty here is that  we need prior knowledge about 
the total time derivative of the velocity as well as the total 
time derivative of the density.
The former one is related to the velocity dissipation.
Once the dissipation formulation is known,
the time evolution of the density can be deduced.

% The dissipation formulation,
% Eq.~\eqref{eq:49} 
% is particularly chosen to 
% be consistent with the Newton viscous stress tensor 
% to dissipate the velocity and the associated energy as 
% \begin{equation}
% \int_\Omega \rho \mathbf{u}\cdot \frac{\mathrm{d}  \mathbf{u}}{\mathrm{d} t}\mathrm{d} \Omega=
% -\tau_1\int_\Omega \nabla (\rho \mathbf{u}): \nabla (\rho \mathbf{u}) \mathrm{d} \Omega
% -\tau_2\int_\Omega [\nabla\cdot (\rho \mathbf{u})]^2 
% \mathrm{d} \Omega\leq0.
% \end{equation}
% Other dissipation formations for $\mathrm{d} \mathbf{u}/\mathrm{d} t$ are also possible as long as the 
% integration 
% \begin{equation}
% \int_\Omega \rho \mathbf{u}\cdot \frac{\mathrm{d}  \mathbf{u}}{\mathrm{d} t}\mathrm{d} \Omega 
% \end{equation}
% is not greater than 0.

\subsection{Remarks} 
\label{sec:13}
The key point of the present theory is the energy minimization based 
on an alternative formulation for the kinetic energy,
$\rho \mathbf{u}\cdot \mathbf{u}$, rather than the conventional formulation,
$\rho \mathbf{u}^2$.
In our derivation, the dissipation of the kinetic energy
is divided into two parts.
The first part is the momentum conservation, $\mathrm{d} (\rho \mathbf{u})/\mathrm{d} t$
which changes not only the fluid velocity $\mathbf{u}$
but also the density $\rho$.
This derivation differs from Euler and Navier-Stokes equations
where the momentum variation is only caused by the evolution
of the fluid velocity.
The second part of the kinetic energy dissipation
is the velocity dissipation associated with the velocity entropy.

In the current work, we consider the linear momentum
$\boldsymbol\zeta=\rho \mathbf{u}$.
Non-linear terms related to fluid velocity
may be accounted for the dissipation if we consider 
non-linear momentum, for example, $\boldsymbol\zeta\propto \mathbf{u}^n$
or $\boldsymbol\zeta\propto (\nabla \mathbf{u})^n$.
The non-linear momentum not only modifies
 the velocity dissipation
 but also changes the momentum balance, which may be applied 
 for understanding highly nonlinear effects.
These specialized cases are out of the scope of the present discussion.

Let us recall the momentum balance equation
\begin{equation}
 \rho\frac{\mathrm{d} \mathbf{u}}{\mathrm{d} t}+\mathbf{u}\frac{\mathrm{d} \rho}{\mathrm{d} t}=-\nabla p.
\end{equation}
The evolution of the velocity corresponds to the dissipation according to Eq.~\eqref{eq17}.
The local velocity can increase (acceleration) or decrease (deceleration), but overall the kinetic energy $\mathcal K$ 
plus other potential energies decreases with time
in a closed system.
Once dissipation takes place, i.e., $\mathrm{d} \mathbf{u}/\mathrm{d} t\neq0$, the density at the interface cannot be assumed to be a constant;
in this case, the density evolves with time.
The density evolution
is especially important when the fluid velocity is 
close to the speed of sound, i.e., high Mach number fluid.
Note that in reality, dissipation cannot be avoided, since entropy always increases with time.

In Newton's law, the time evolution of the velocity is defined as the ``acceleration'',
where the density evolution within the interface has been overlooked.
Our concept provides a new understanding on the ``acceleration''.
The associated time evolution of the velocity corresponds to certain types of 
dissipation, coupled with the density evolution, which leads to the momentum balance.
More generally speaking, we have divided the kinetic energy evolution into two parts:
\begin{equation}
 \frac{\mathrm{d} \mathcal K}{\mathrm{d} t}
 =\int_\Omega \bigg[\frac{\mathrm{d}  (\rho \mathbf{ u}
 )}{\mathrm{d} t}\cdot \mathbf{u}+\rho \mathbf{ u}\cdot \frac{\mathrm{d}  \mathbf{u}}{\mathrm{d} t} \bigg]\mathrm{d} \Omega.
\end{equation}
The first part corresponds to 
the momentum balance. The second part is 
associated with the velocity dissipation.
% Overall, $\mathrm{d} \mathcal K/\mathrm{d} t\leq0$, which is the key basis of the present finding.

\section{Case Studies and Applications: Validation and Broader Implications}
\label{sec:case}

In this section, we apply the present theory to three representative cases:
(A) sound propagation in air, (B) equation of state (EOS) for ideal gas, 
and (C) the motion of a water-air interface in a tube,
to demonstrate the validity of the proposed theory.
In case A, we derive a general formulation for the sound speed that 
is consistent with previous perturbation analysis. 
In case B, we show that the present density evolution equation in steady state
generalizes the EOS of ideal gas.
In case C, we demonstrate the validity of Bernoulli's law, 
as well as its generalization accounting for the density jump at the water-air interface.

\subsection{Sound speed}

When a sound wave propagates through air, 
the local density of the air varies with time—a hallmark of compressible flow. 
In an ideal gas, sound transmission occurs through molecular collisions. Over the timescale of these collisions, entropy remains constant (isentropic condition), implying no dissipation in the sound speed $c$ (i.e., $\mathrm{d} c/\mathrm{d} t=0$).
Consequently, momentum evolution arises solely from density variations at the wavefront:
\begin{equation}
 c\frac{\mathrm{d} \rho}{\mathrm{d} t}=-\frac{\mathrm{d} p}{\mathrm{d} x} -\eta \frac{\mathrm{d}^2c}{\mathrm{d}x^2}.
 \label{eq:81}
\end{equation}
We postulate that sound propagation in a gas is driven by local variations in air density, not by changes in fluid velocity. Under the condition $c=\mathrm{d} x/\mathrm{d} t$, we derive the following equation at the wavefront 
when assuming isentropic condition ($\eta\approx0$) via omitting the viscous term:
\begin{equation}
 c^2=-\frac{\mathrm{d}p}{\mathrm{d}\rho}.
\end{equation}
Using the pressure definition, $-p=f-\sum_i\mu_i \phi_i$, we derive a general expression for the sound speed:
\begin{equation}
 c^2=\frac{\mathrm{d} (f-\sum_i\mu_i \phi_i)}{\mathrm{d} \rho},\ \text{or},\ c=\sqrt{\frac{\mathrm{d} (f-\sum_i\mu_i \phi_i)}{\mathrm{d} \rho}}=\sqrt{\frac{\mathrm{d} \Psi}{\mathrm{d} \rho}},
 \label{eq:83}
\end{equation}
which is consistent with the results of the perturbation analysis~\cite{landau2013fluid}.
Note that two conditions have been applied 
to derive the sound speed:
$\delta q\approx0$  and $\eta\approx0$, known as isentropic condition (adiabatic and reversible).
For one component fluid, the Landau potential is expressed as $f-\sum_i\mu_i \phi_i=f$,
where the free energy $f$ is written as $f=\theta-Ts$.
In the case of isentropic condition, the free energy is mainly contributed by the internal energy, which consists
of microscopic kinetic energy $\varrho$ and microscopic potential energy $\xi$, 
\begin{equation}
 \theta=\varrho+\xi.
\end{equation}
The microscopic kinetic energy arising from the thermal motion of the gas molecules is expressed
as  $\varrho=\rho \bar v^2$,
where $\bar v$ is the mean thermal velocity of the gas  molecules.
Performing its derivative to the density according to Eq.~\eqref{eq:83}, we obtain the sound speed as 
\begin{equation}
 c=\sqrt{\bar v^2+\frac{\mathrm{d}\xi}{\mathrm{d}\rho}}.
\end{equation}
For ideal gas, there is no interaction between gas molecules, and there is only the microscopic
kinetic energy, namely,   $\xi=0$, resulting in 
\begin{equation}
 c\approx \bar v.
\end{equation}
This leads to an important conclusion 
that the sound speed in an ideal gas can be approximated by the mean thermal
 motion velocity of the gas molecules.
 For example,
 in air, the sound speed at room temperature and 1 bar is  about 340 m/s,
 which is quite close to mean thermal velocity of gas molecules estimated 
 via $\sqrt{R_gT/M}\approx 300$ m/s ($R_g$: universal gas constant, $M$: molecular weight).
 It must be noted that the above calculation only considers the kinetic energy
 and overlooks the interaction energy between gas molecules.
 The latter one cannot be neglected in real gas, solid, and liquid.

We further demonstrate that 
 the density evolution equation is also consistent with the
 dispersion relation of sound propagation. 
When a sound wave propagates through air, the local air density varies temporally, which may be expressed as:
\begin{equation}
 \rho(x,t)=\rho_0\exp[-i(kx-\omega t)],
\end{equation}
where $k$ represents the wave number, $\omega$ depicts the frequency,
and $\rho_0$ is a constant.
Substituting the density wave into Eq.~\eqref{eq:81}, 
the density evolution equation replicates
the dispersion relation of sound wave 
\begin{equation}
 \lambda= 2\pi c/\omega.
\end{equation}
Here, $\lambda=2\pi/k$ is the wavelength of sound speed.

\subsection{Equation of the state (EOS)}
% In the second example, we demonstrate that the current density evolution equation reproduces and generalizes the ideal gas equation of state (EOS).

In the steady state, the density evolution for inviscid fluids reduces to 
\begin{equation}
 \rho u^2\frac{1}{\rho}\frac{\mathrm{d} \rho}{\mathrm{d} x}=-\frac{\mathrm{d} p}{\mathrm{d} x},
 \label{eq:100}
\end{equation}
where we have added a constant of unity: $\frac{\rho}{\rho}=1$, to the left hand side.
When comprehending $u$ as the microscopic velocity,
the kinetic energy $\rho u^2$
should equal to the thermal motion energy, $R_gT/v_m$ for ideal gas ($v_m$: molar volume).
Integrating from a reference state 
with density $\rho_0$ and pressure $p_0$
to the  state to be solved,
Eq.~\eqref{eq:100} is rewritten as 
\begin{equation}
 \frac{R_gT}{v_m}\int_{\rho_0}^\rho \frac{1}{\rho}\mathrm{d}\rho=-\int_{p_0}^p \mathrm{d}p.
\end{equation}
Replacing $v_m$ by $V/n$ ($V$: volume of the gas, $n$: number of mole), we obtain the following  EOS 
\begin{equation}
 nR_gT(\ln\rho-\ln \rho_0)=(p_0-p)V.
\end{equation}
When choosing $p_0=0$ when $\ln \rho_0=0$, we obtain 
\begin{equation}
 nR_gT\ln\rho=-pV=(f-\sum_i\mu_i\phi_i)V=\Psi V,
 \label{eq:92-00}
\end{equation}
which is the EOS of ideal gas with a factor, $\ln\rho$.
Note that the thermodynamic pressure $\Psi$ should be used instead of $p$.
The term $\ln\rho$ is analogous to the entropy contribution, and this result is consistent with previous work on the free energy formulation of a water-gas system~\cite{ledesma2014lattice}.
By dividing both sides of Eq.~\eqref{eq:92-00} by the molar volume, we obtain $\Psi \sim \rho \ln \rho$,
which is the prescribed EOS used in Refs.~\cite{ledesma2014lattice,ridl2018lattice}.
This highlights our central result: the density evolution in our theory naturally generates EOS, 
in contrast to a prescribed EOS  in existing diffuse-interface methods.

\subsection{Water-air interface}
As a further application, we consider a water-air system to demonstrate the present theory. 
Following the van der Waals-Cahn diffuse interface approach, the free energy functional for a binary system 
takes the form:
\begin{equation}
 \mathcal F=\int_\Omega  
 \bigg[\frac{\gamma}{\epsilon}\phi^2(1-\phi)^2 +\gamma\epsilon(\nabla \phi)^2 \bigg]\mathrm{d}\Omega.
\end{equation}
The first term is a double-well potential with two local minima corresponding to the bulk water and bulk air phases.
Interpreting $\phi$ as the volume concentration of water,
 $\phi=1$ represents pure water phase
and $\phi=0$ corresponds to the pure gas phase.
The second term accounts for the gradient energy density, originating from a Taylor expansion of the energy associated with concentration gradients at the interface~\cite{cai2024chemo}.
The parameters $\gamma$ and $\epsilon$
control the interfacial tension and width, respectively. 
For simplicity, we set $\gamma=1$ and $\epsilon=1$ without loss of generality.

\begin{figure}[h!]
 \centering
 \includegraphics[width=0.56\linewidth]{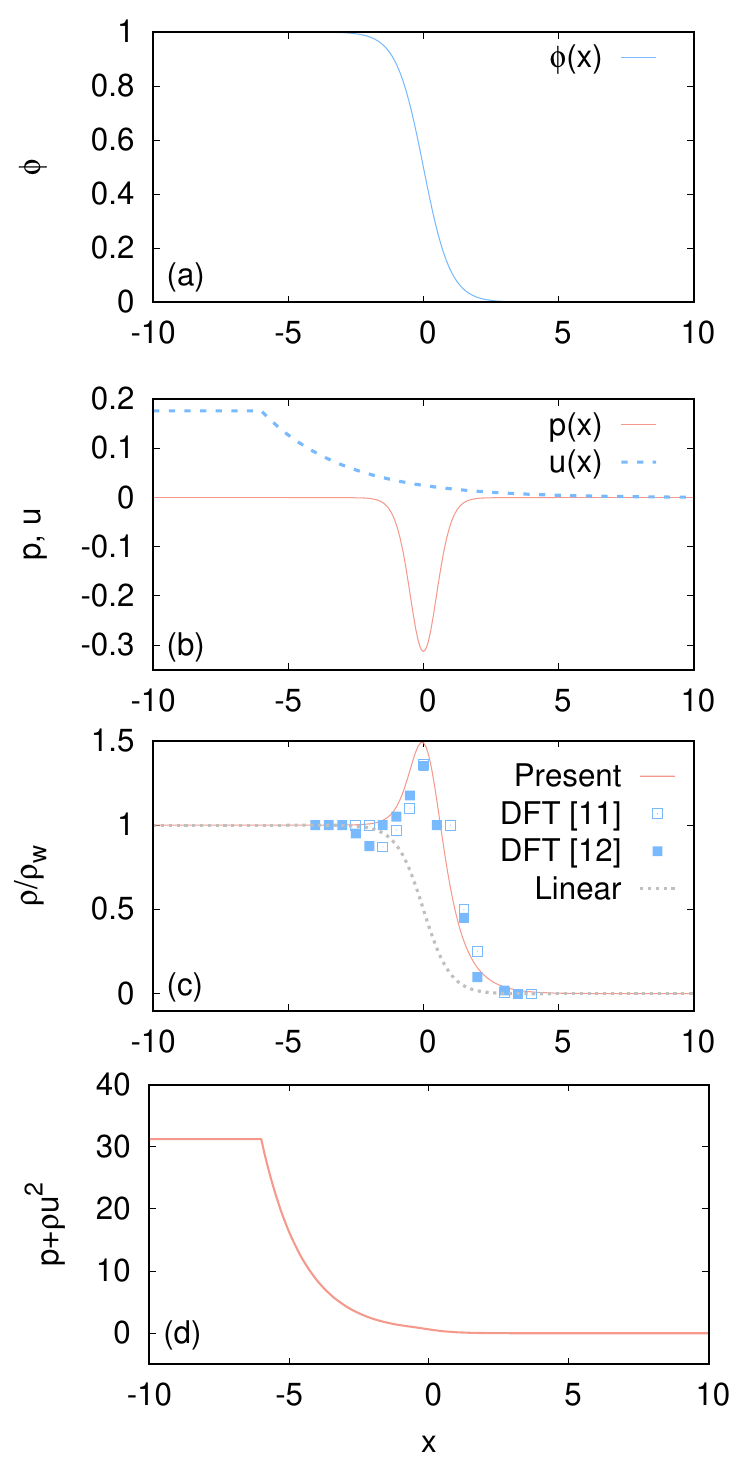}
 \caption{Analytical solution of the equation system, Eq.~\eqref{eq:94}-Eq.~\eqref{eq:98-00} in 2D for a moving flat water-air interface in a tube:
 (a) Volume concentration $\phi(x)$. (b) Pressure $p(x)$ and fluid velocity $u(x)$, with $u_0=0.025$ and $a=0.33$.
 (c) Density $\rho(x)/\rho_w$, where $\rho_w$ is the density of bulk water.
 Square symbols represent DFT results from   Refs.~\cite{pezzotti20182d,creazzo2024water};
the dashed line shows a linear interpolation of the density.
 (d) Sum of the pressure energy and  kinetic energy, $p+\rho u^2$.}
 \label{fig5}
\end{figure}

To minimize the free energy plus pressure and kinetic energies, 
the time evolution equations of the system read
\begin{align}
\label{eq:94}
&\frac{\partial \phi}{\partial t}+\mathbf u \cdot \nabla\phi =\nabla\cdot \varsigma \nabla \mu,\\
&\mu =2\phi(1-\phi)(1-2\phi)-2\nabla^2\phi,\\
&\mathbf{u}\frac{\mathrm{d} \rho}{\mathrm{d} t}=\mathbf{u}\bigg(\frac{\partial \rho}{\partial t}+\mathbf{u}\cdot \nabla \rho \bigg)=-\nabla p,\\
&\frac{\mathrm{d} \mathbf{u}}{\mathrm{d} t}=\frac{\partial \mathbf{u}}{\partial t}+\mathbf{u} \cdot \nabla \mathbf{u}=-\tau_0 \mathbf{u},\\
&p= -\phi^2(1-\phi)^2 -(\nabla \phi)^2+\mu \phi.
\label{eq:98-00}
\end{align}
The system comprises five equations for five unknowns: $(\phi, \mu, \rho, \mathbf{u}, p)$.
With appropriate boundary conditions, it admits a unique solution.
In the following, we derive an analytical solution in 2D for a flat water-air interface moving towards the gas
inside a tube with full slip condition (i.e.,
no wall friction, corresponding to a Young's contact angle of 90$^o$).

The analytical solution reads
\begin{align}
&\phi(x)=\frac{1}{2}[1-\tanh(x)],\\
&\mu(x)=0,\\
&u^2\frac{\mathrm{d} \rho}{\mathrm{d}  x}=-\frac{\mathrm{d} p}{\mathrm{d} x}, \label{eq:98} \\
&u(x)=u_0\exp(-ax), \label{eq:102}\\
&p(x)=-\phi^2(1-\phi)^2-\frac{1}{4}[1-\tanh^2(x)]^2.
\end{align}
The solution fulfils the boundary conditions
\begin{itemize}
 \item [(B1)] $\phi=1$ when $x\rightarrow -\infty$; $\phi=0$ when $x\rightarrow +\infty$.
 \item [(B2)] $\frac{\mathrm{d} \phi}{\mathrm{d} x}=0$ when $x\rightarrow -\infty$ and  $x\rightarrow +\infty$,
 i.e., no gradient of composition in the bulk water and gas.
 \item [(B3)] According to (B1) and (B2), the boundary condition of the chemical potential is 0, i.e.,
 $\mu(x)=0$ everywhere.
 \item [(B4)] The fluid velocity in the water far from the interface is a constant, $u_w=u_0$.
 In the gas far from the interface,  the fluid velocity is zero,  $u_a=0$.
 \item [(B5)] The bulk pressure of the water ($p_w$)
 and the bulk pressure of the gas ($p_a$) both are the same, 
 fulling the Young-Laplace relation for a flat water-air interface.
\end{itemize}

The above solution is illustrated in Fig.~\ref{fig5} 
and indicates the 
following results:
\begin{itemize}
 \item [(i)] In the bulk water and bulk gas phases, we have
 $p_w+\frac{1}{2}\rho_wu_w^2=C_1$
 and  $p_a+\frac{1}{2}\rho_gu_a^2=C_2$, where $C_1$ and $C_2$ are two distinct constants.
 These results validate Bernoulli's law.
 The difference in these two constants is  proportional to the density difference $(\rho_w-\rho_a)$.
 This energy jump has not been captured
 by Bernoulli's law. 
  \item [(ii)] The pressure across the water-air interface first decreases from a constant value 
 $p_w=0$ until the position $\phi (x=0)=0.5$
  and then increases to a constant value  $p_a=0$ in the gas phase.
  The space distribution is symmetric for a flat interface.
  For a curved water-air interface, Eq.~\eqref{eq:94}
  reproduces the Young-Laplace equation,
  \begin{equation}
   \Psi_w-\Psi_a=\sigma_{wa}\varGamma,
   \end{equation}
  which has been already discussed in Ref.~\cite{wang2023thermodynamically} ($\sigma_{wa}$: surface tension; $\varGamma$: mean curvature).  
  \item [(iii)] The fluid velocity decays from a constant value in the bulk water
  to zero far from the interface in the gas.
  The spatial decay is linear when $\tau_0$ is a positive constant, as demonstrated by the steady state equation for fluid velocity:
  \begin{equation}
   u\frac{\mathrm{d} u}{\mathrm{d} x}=-\tau_0 u.
  \end{equation}
  As given by Eq.~\eqref{eq:102}, the decay is exponential when $\tau_0$ is proportional to $u$, namely, $\tau_0=au$,
  where $a$ is a positive constant.
The decay can also be Newtonian type
\begin{equation}
 u\frac{\mathrm{d} u}{\mathrm{d} x}=\eta \frac{\mathrm{d} ^2u}{\mathrm{d} x^2},
\end{equation}
which will not be discussed in this work.  
 \item [(iv)] The density profile across the water-air interface is 
 obtained via solving the ordinary differential equation, Eq.~\eqref{eq:98}
 with the boundary condition $\rho=\rho_w$ when $x\rightarrow -\infty$
 and $\rho=\rho_a$ when $x\rightarrow +\infty$.
 Due to the special property of the pressure
 that the derivative $-\mathrm{d} p/\mathrm{d} x$ is positive
 for $x\in(-\infty, 0)$
 and the derivative $-\mathrm{d} p/\mathrm{d} x$ is negative
 for $x\in(0, \infty)$, 
 the density first increases from $\rho=\rho_w$ to a maximum value at $x=0$
 and then decreases to $\rho=\rho_a$ in the gas phase.
 The occurrence of a density peak 
 within the water-air interface is fully consistent with recent results of the
 density functional theory (DFT)~\cite{pezzotti20182d,creazzo2020enhanced,creazzo2024water}.
 A change in the formulation of 
 the velocity decay cannot remove the peak,
 but can affect the peak value
 as well as the peak width. 
 It must be noted the conventional interpolation approach,
 $\rho=\rho_w\phi+\rho_a(1-\phi)$ can only lead
 to a monotonic decrease of the density from $\rho_w$ to $\rho_a$, contradicting
 with the DFT results.  
\end{itemize}

\section{Conclusion}
\label{sec:14}
In summary, we have proposed an 
alternative theory, distinct from the Navier-Stokes 
and Euler equations, to address the high-density-ratio problem in fluid dynamics.
Our approach is grounded in the fundamental principle of energy minimization. 
A noteworthy outcome of this work is the derivation of a novel density evolution equation
\begin{equation}
 \mathbf u\dot \rho
 =-\nabla \cdot [(p\mathbf{I}+\underline{\underline{\boldsymbol \Theta}})]
 -\rho\nabla \cdot [ \tau_1(\nabla  \boldsymbol\zeta+\nabla\boldsymbol\zeta^T)+\tau_2 (\nabla\cdot \boldsymbol\zeta)\mathbf{I}] 
\end{equation}
for the high density ratio system, such as water-air.
From this equation, we see that
the density evolution or the local excess volume is affected by two factors:
\begin{itemize}
 \item the local pressure $p$ and stress tensor $\underline{\underline{\boldsymbol \Theta}}$;
 \item the viscous dissipation.
\end{itemize}
The latter one provides a supplement to the Navier-Stokes
equation that the density can be altered by the viscous dissipation.
The validity of our density evolution equation has been 
demonstrated for inviscid fluids. Specially, our derivation leads
to a generalization of Bernoulli's law.
In addition, we have shown
that the current density evolution is consistent with 
\begin{itemize}
 \item sound speed in fluids;
 \item dispersion relation of sound wave;
 \item EOS of ideal gas;
 \item DFT results for 
 non-monotonic density across the water-air interface.
 \end{itemize}
We expect that the present model can
be used for immiscible fluids with arbitrary density ratios.

\begin{acknowledgements}
We express our sincere gratitude to the Editor, Dr. Cottin-Bizonne, for facilitating constructive exchanges with the referees, and to all referees for their careful and insightful comments.
\end{acknowledgements}

\appendix

\section{The modified Gibbs-Duhem equation and the pressure}

The modified Gibbs-Duhem relation reads
\begin{equation}
 -\delta p + \sum_{i=1}^K \phi_i\delta \mu_i+ \rho \mathbf{u}  \cdot \delta\mathbf{u}  =0.
 \label{eq:a1}
\end{equation}
The velocity term is replaced by the entropy contribution as  $ \rho\mathbf{u} \cdot \delta \mathbf{u}=-T\delta s_u=-\delta (Ts_u)$.
In the last equality, we have used the isothermal condition.
The chemical potential term $\phi_i\delta \mu_i$ is rewritten as 
$\phi_i\delta \mu_i=\delta (\mu_i \phi_i)-\mu_i \delta \phi_i$ according to the chain rule of total derivative.
Recalling the definition of the chemical potential, $\mu_i=\delta e/\delta \phi_i$,
we further write the chemical potential term $\phi_i\delta \mu_i$ 
as $\sum_{i=1}^K\phi_i\delta \mu_i=\sum_{i=1}^K\delta (\mu_i \phi_i)-\delta e$ (see Appendix C).
Summarizing all these equalities,
we obtain the following equation 
\begin{equation}
 \delta (-p+\sum_{i=1}^K\mu_i \phi_i-e-Ts_u)=\delta (-p+\sum_{i=1}^K\mu_i \phi_i-e_0)=0.
 \label{eq:a2}
\end{equation}
In the above equation, we have used the equality $e+Ts_u=e_0$.
Integrating Eq.~\eqref{eq:a2}, we obtain the final expression for the pressure as 
\begin{equation}
 -p=e_0-\sum_{i=1}^K\mu_i \phi_i+p_\text{ref},
 \label{eq:a4}
\end{equation}
where $p_\text{ref}$ is a constant.
For simplicity, this constant is set as zero in the current work.

\section{ $\delta$ operator,  $\nabla$ operator, and the formulation of the  chemical potential}
If replacing the $\delta$ operator by the space gradient operator $\nabla$
and the velocity term by the velocity entropy $-T\nabla s_u$
in Eq.~\eqref{eq:a1},
the following equation for the gradient of the pressure is obtained 
\begin{equation}
 -\nabla p =- \sum_{i=1}^K \phi_i \nabla \mu_i+T\nabla s_u.
 \label{eq:a3}
\end{equation}
 
When using the original definition of the  chemical potential as
\begin{equation}
 \mu_i=\frac{\mathrm{d}  e}{\mathrm{d}\phi_i}= \partial e/\partial \phi_i+ 
(\partial e/\partial \nabla \phi_i)\cdot \nabla\phi_i,
 \label{eq:b2}
\end{equation}
which is consistent with Ref.~\cite{ridl2018lattice},
Eq.~\eqref{eq:a3} is rewritten as 
\begin{equation}
 -\nabla p-\nabla (T s_u)+\sum_{i=1}^K [\nabla(\mu_i \phi_i)-\mu_i \nabla \phi_i]=0.
  \label{eq:b3}
\end{equation}
Integrating Eq.~\eqref{eq:b3} by using Eq.~\eqref{eq:b2},
we obtain the same formulation of Eq.~\eqref{eq:a4}.

Note that if the conventional functional derivative
\begin{equation}
 \mu_i= \partial e/\partial \phi_i-\kappa_i\nabla^2\phi_i=\partial f_0/\partial \phi_i-\kappa_i\nabla^2\phi_i
 \label{eq:b0}
\end{equation}
is used~\cite{ledesma2014lattice}, one cannot obtain  Eq.~\eqref{eq:a4}.
By using  Eq.~\eqref{eq:b0},
one obtains a different formulation of the pressure $p=f_0-\sum_{i=1}^K \kappa_i (\nabla \phi_i)^2- \sum_{i=1}^K \mu_i\phi_i
+p_\text{ref}$; a negative sign occurs in the second term.
The reason for this inconsistency is as follows.
Only after applying the integration by parts 
to the term $ (\partial e/\partial \nabla \phi_i)\cdot \nabla\phi_i$  in Eq.~\eqref{eq:b2} 
on the whole domain $\Omega$,
 we obtain the second term with a negative sign in Eq.~\eqref{eq:b0}.
 This negative sign arises from integration by parts and leads to inconsistent results for the pressure.
 One cannot use this result—obtained after applying integration by parts in the whole domain—to integrate the Gibbs–Duhem relation,
 which is a local differential equation.
 Instead, the original definition given in Eq.~\eqref{eq:b2} should be used when integrating the Gibbs–Duhem relation.
 
\section{Derivative of $\mu \phi$}
Supposing that $\phi(\mathbf x)$ and $\phi(\mathbf y)$ are functions: $\mathbb R^3\mapsto\mathbb R$,
and $\mu$ is a functional: $\Phi \mapsto\mathbb R$, where $\phi(\mathbf  x),\ \phi(\mathbf  y)\in\Phi $,
the derivative for the product of $\mu$ and $\phi$ reads~\cite{gelfand2000calculus}
\begin{equation}
 \frac{\delta [\mu \phi(\mathbf  x)]}{\delta \phi(\mathbf  y)}=\phi(\mathbf  x) \frac{\delta \mu }{\delta \phi(\mathbf  y)}+\mu  \frac{\delta  \phi(\mathbf  x)}{\delta \phi(\mathbf  y)}.
\end{equation}
In a special case, $\phi(\mathbf  x)=\phi(\mathbf  y)$, we have $\frac{\delta  \phi(\mathbf  x)}{\delta \phi(\mathbf  y)}=1$, and therefore
$\delta (\mu \phi)=\phi \delta \mu+ \mu \delta \phi$.
As $\mu=\delta e/\delta \phi$,
the above equation is finally expressed as 
$\delta (\mu \phi)=\phi \delta \mu+  \delta e$.

\section{The problem of the classic continuity equation for compressible fluids}
For one-component fluids, the mass conservation reads
\begin{equation}
  \frac{\partial }{\partial t}\int_V nm_0 \mathrm{d}V=-\int_\Gamma \mathbf{j}\cdot \mathbf{n}\mathrm{d}\Gamma,
\end{equation}
where $n$ is the number of the mass particles, such as molecules or atoms,
and $m_0$ is the  mass per particle.
The flux $\mathbf{j}$ represents the flow of mass particles across the boundary $\Gamma$,
with normal vector $\mathbf{n}$.
By applying  Gauss's divergence theorem, we obtain 
\begin{equation}
 \frac{\partial (nm_0)}{\partial t}+\nabla \cdot \mathbf{j}=0.
 \label{eq:d2}
\end{equation}
To convert Eq.~\eqref{eq:d2} into 
an evolution equation for the density, both sides must be divided by the volume element $V$.
Based on the following calculation for $V$:
\begin{equation}
 V=v_m+v_e,
\end{equation}
where $v_m$ is the volume of molecules or atoms,
and $v_e$ is the excess volume which can also be interpreted as the vacancy volume,
we classify the discussion into two cases: 
(I) When the vacancy volume is much smaller than the molecular volume, i.e. i.e., $v_e\ll v_m$, as in a one-component solid or a solid-like liquid, the system corresponds to a nearly incompressible phase.
(II) When the vacancy volume is comparable to the molecular volume, as in a one-component gas, the system corresponds to a compressible fluid.
In case (I), the system cannot be significantly compressed, and the local volume element $V$ can be considered nearly constant.
By defining the flux $\mathbf j=m_0n\mathbf{u}$ and multiplying both sides by the approximately constant factor 
$1/V\approx 1/v_m$,
we recover the classic continuity equation
\begin{equation}
\frac{\partial \rho }{\partial t}+\nabla \cdot (\rho \mathbf u)=0.
\label{eq:D4}
\end{equation}
In case (II), the local volume element $V$ can vary with time and space. Consequently, the continuity equation,
 Eq.~\eqref{eq:D4} takes the form:
\begin{equation}
 (nm_0)\frac{\partial (1/V) }{\partial t}+\frac{1}{V}\frac{\partial (n m_0) }{\partial t}+\nabla \cdot (\frac{nm_0}{V} \mathbf u)=0.
\label{eq:D5}
 \end{equation}
Due to the large amount of vacancies in case II,  $(1/V)$ is not guaranteed to be a conserved variable, since vacancies can appear or disappear locally under the influence of pressure.
As such,  conventional models require a prescribed equation of state (EOS) to define 
$1/V$,
for example, $pV=\text{constant}$ for an ideal gas.
Moreover, when  $(1/V)$ is not conserved,
the convection flux cannot be formulated as $\mathbf{j}=\rho \mathbf{u}$,
because the density $\rho$ contains information from $1/V$. Incorporating a non-conserved variable into the mass conservation equation is therefore not appropriate.

The same issue arises for multicomponent compressible fluids with $K\in \mathbb N$ components, where $(1/V)$ is also  not conserved. The only difference is that the  volume element is rewritten as $V=\sum_{i=1}^K v_i +v_e$,
where $v_i$ is the volume of the $i$-th component.

In the static state with $\mathbf{u}=\mathbf{0}$, Eq.~\eqref{eq:D4} or Eq.~\eqref{eq:D5}
returns to a trivial equality: $0=0$, which has no physical meaning.
In this case, we have to consider the microscopic velocity to derive EOS (Section VII B).
Within a moving interface indicating that $1/V$ is non-uniform in space,
the continuity equation Eq.~\eqref{eq:D4} cannot be applied,
as $\partial_t (1/V)+\nabla \cdot (\mathbf{u}/V)$ is not necessary to be zero,
since the vacancy can vanish or appear locally in the moving process of the interface.
% Even when the velocity $\mathbf{u}\neq\mathbf{0}$ is uniform within the interface,  at the steady state ($\partial_t (1/V)=0$), 
% there can still be a nonzero gradient of the volume element within the interface, i.e.,
%  $\mathbf{u}\cdot\nabla(1/V)\neq0$.

In our model, we consider two distinct evolution equations.
The first is for the volume concentration $\phi$ including the excess volume and the volumes of all other mass species.
The evolution of $1/V$ is driven by the maximization of the mixing entropy $s_\phi$, resulting in a diffusion equation.
For single-component fluids, this diffusion occurs via the exchange of molecules with vacancies, such as in self-diffusion.
For multicomponent fluids, diffusion occurs through the exchange of molecules with both vacancies and other molecular species.
The second evolution equation governs the density. For the approximation of inviscid fluids, 
the density evolution in our model based on Eq.~\eqref{eq23} is expressed as
\begin{equation}
 \mathbf{u}\frac{ \mathrm{d} \rho}{ \mathrm{d} t}=\mathbf{u}\bigg(\frac{\partial \rho}{\partial t}+\mathbf{u}\cdot \nabla \rho\bigg)=-\nabla p.
  \label{eq:d5}
\end{equation}
A heuristic understanding for Eq.~\eqref{eq:d5} is that $\nabla p \propto \nabla (1/V)$ under a prescribed EOS,
which accounts for local compression effects on the volume element and thereby extends the classical continuity equation.
% This understanding roughly explains how our model can naturally recover the equation of state (see Section VII B).
For incompressible fluids, the pressure work is nearly zero, $pdV\approx 0$,
and therefore $-\nabla p\approx 0$ in the Gibbs-Duhem relation.
In this case, Eq.~\eqref{eq:d5} reduces to the classic continuity equation for incompressible flow.
This scenario corresponds to 
$v_e=0$ i.e., a rigorously incompressible fluid with no vacancies.
For multicomponent incompressible fluids, the density evolution of each component $\rho_i$ can be solved separately.
For compressible fluids ($v_e \neq 0$), the density evolution of individual components cannot generally be solved separately,
since the effect of $v_e$ on $v_1$, $v_2$, $\cdots$, $v_K$
is unknown. In our model for compressible fluids,  Eq.~\eqref{eq:d5} naturally gives rise to  the equation of state (Section VII B), so a prescribed EOS is not required.  However, if a prescribed EOS is available~\cite{ridl2018lattice}, which implicitly prescribes the effect of $v_e$, then the density evolution of different components can, of course, be solved separately.

\end{document}